\documentclass[aps,pre,reprint,showpacs]{revtex4-1}

\usepackage{amsmath}
\usepackage{graphicx}% Include figure files
\usepackage{dcolumn}% Align table columns on decimal point
\usepackage{bm}% bold math
\usepackage{braket}
\usepackage{subfigure}

\newcommand\Vmax{V_{\text{max}}}

\begin{document}

\preprint{Ver 5.2 - Mar 15}

\title{Finite Size Effects in the Nagel-Schreckenberg Traffic Model}

\author{Ashkan Balouchi}
\email{balouchi.ashkan@gmail.com}
\author{Dana A. Browne}
\email{phowne@lsu.edu}
\affiliation{
Department of Physics and Astronomy,
Louisiana State University, Baton Rouge, LA 70803}

\date{\today}

\begin{abstract}

We examine the Nagel-Schreckenberg traffic model for a variety of maximum
speeds.  We show that the low density limit can be described as a dilute
gas of vehicles with a repulsive core.  At the transition to jamming,
we observe finite-size effects in a variety of quantities describing the
flow and the density correlations, but only if the maximum speed $\Vmax$
is larger than a certain value.  A finite-size scaling analysis of several
order parameters shows universal behavior, with scaling exponents that
depend on $\Vmax$.  The jamming transition at large $\Vmax$ can be viewed
as the nucleation of jams in a background of freely flowing vehicles.
For small $\Vmax$ no such clean separation into jammed and free vehicles
is possible.

\end{abstract}

\pacs{45.70.Vn,89.40.Bb}

\maketitle

\section{Introduction\label{intro}}

The flow of traffic represents a many-particle non-equilibrium problem with
important practical consequences.  Traffic flow shows well defined
collective behavior where the free flow of traffic at low density
changes abruptly with growing density to a denser phase with jams.
The jams themselves show organized motions with start-stop waves as the
cars creep forward.  In addition to free flow and jam phases, there are
also instances of synchronized flow at low velocity.  Understanding the
collective dynamical behavior and controlling the jams will give insight
into effective traffic management.  The flow of traffic also gives us an
example of a non-equilibrium system with a phase transition, and it is
interesting to inquire into the cause of the transition and whether
long-range correlations appear near the transition.

Traffic behavior has been studied for decades, using a variety of
approaches, including fluid dynamics models~\cite{FluidDyn}, Boltzmann
equations~\cite{Boltz}, and most recently, cellular automaton (CA)
approaches~\cite{Chowdhury2000,Helbing2001,Nagatani2002}.  In CA models,
the vehicles occupy discrete sites and have discrete velocities, hopping
from site to site according to simple rules.  Despite their simplicity,
these CA models appear to capture much of the collective behavior observed
in real traffic.

In 1992, Nagel and Schreckenberg (NS)~\cite{Nagel1992} introduced a
relatively simple CA model for traffic flow.  The road is represented
as a set of $L$ equally spaced sites, each of which can be occupied by
at most one of $N$ vehicles.  Vehicles have discrete velocities $v_{n}$
from $0$ to a maximum velocity $\Vmax$.  The hopping dynamics follow 4
simple rules, applied in the following sequence. First, each car with
$v_{n}<\Vmax $ increases its velocity by one.  The gap to the next car
$g_{n}=r_{{n+1}}-r_{n}-1$ is computed, where $r_n$ is the position of
the n-th car.  If the car has a speed greater than $g_{n}$, it will
brake to reduce its speed to $g_{n}$ to prevent a collision (gap rule).
Variations in driver behavior are modeled by then lowering the speed
$v_{n}\to v_{n}-1$ with a fixed probability $p$.  Finally, the position
is updated via $r_{n}\to r_{n}+v_{n}$.

The NS model mimics some, but by no means all, of the observed features
of traffic flow.  As the first of a series of increasingly detailed CA
traffic models, it has been widely studied during the past twenty years to
understand the nature of the phase transition from free flow to jams.
Nagel and Schreckenberg~\cite{Nagel1992} showed that at low density
$d=N/L$, a free flow state occurred where the cars all have a speed of
$\Vmax$ or $\Vmax-1$, with a mean speed of $\Vmax-p$.  At a certain
density, the steady state changes to a phase with a nonzero fraction
of the cars participate in a jam of slowly moving or stopped vehicles.
Nagel and Paczuski~\cite{Nagel1995} showed in a variant of the NS
model, where cars with $V=\Vmax$ maintain their velocity as a kind of
cruise control, that the jam lifetime showed a power law distribution
at the transition to the jam phase.  Lubeck et al.~\cite{Lubeck1998}
studied the density distribution in the NS model and suggested that the
free flow and jam phases coexist after the transition.  Chowdhury et
al.~\cite{Chowdhury1997,Chowdhury1998} examined the gap distribution and
time-headway distribution (the time delay between two consecutive cars
passing a site) and also concluded that there is a two phase coexistence
after the transition.  Roters et al.~\cite{Roters1999} investigated
the dynamical structure factor and concluded that a continuous phase
transition occurs, but later work~\cite{Chowdhury20002,Roters2000}
suggested that the simulations were not long enough and that the
critical behavior was actually a crossover phenomenon.  Kerner et
al.~\cite{Kerner2002} observed evidence of two first order phase
transitions, with an intermediate phase of synchronized flow between
the free flow phase and the jam phase.

Many quantities have been used to study the transition to the jam
phase.  A number of them use the velocity distribution, such as the
number of stopped cars ($V=0$)~\cite{Vilar1994}, slowly moving cars
($V \le \Vmax/2$)~\cite{Jost2003} or cars not moving at the speed limit
($V<\Vmax$)~\cite{Nagel1995}.  Other authors have chosen the number of
vehicles forced to brake~\cite{Miedema2014} or the difference between the
average velocity and the free flow velocity~\cite{Souza2009}.  All of
these resemble order parameters, being nearly to zero in the free flow
phase and nonzero in the jam phase.  Other quantities have been studied
that not necessarily zero in the free phase, but show an abrupt change at
the transition, such as the vehicle flux~\cite{Nagel1992} or the change
in the vehicle's kinetic energy per step~\cite{Zhang2011,Zhang2014}.
 A number of different traffic
correlations have also been studied, including different characteristic
velocities in the spatial dynamical structure factor~\cite{Lubeck1998},
different maxima in the velocity-position correlation~\cite{Roters1999},
the gap or time-headway distribution~\cite{Chowdhury1998}, the number
of cars moving cooperatively~\cite{DeWijn2012} and velocity correlations
among the cars~\cite{Lakouari2014}.

Despite this effort, a comprehensive picture of the transition in the NS
model is still
incomplete, with different approaches producing differing conclusions
about the nature of the transition or the presence of long range order.
In this paper, we will examine how the value of $\Vmax$ affects the
transition.  We will show that, while the static structure factor shows
long range behavior appearing at the transition for any $\Vmax$, we
only see finite-size effects in the order parameter for $\Vmax \agt 6$.
This indicates that the nature of the long range behavior is different
at high and low values of $\Vmax$.  We use these finite-size effects
to extract the scaling behavior at the transition for several order
parameters.  Our work indicates that the onset of the jam phase can be
analyzed as a two phase coexistence of free flow and localized jams,
as others have observed~\cite{Lubeck1998,Chowdhury1998,Roters1999}.
We show that the presence or absence of long range correlations
can be attributed to a qualitative change in the way jams nucleate
at high and low values of $\Vmax$.

Despite the fact that the NS model misses some features observed
in actual traffic flow~\cite{Kerner2013,Kerner3phase}, it remains a
useful model to probe the nature of a nonequilibrium phase transition.
It also provides a backdrop to understand the transitions observed in
real traffic flow, which exhibits an intermediate synchronized flow
phase~\cite{Kerner2002,Kerner3phase} in some conditions.

In section II of this paper, we review the details of our simulation
and the quantities we use in our analysis.   Section III presents a
quantitative analytic model of the behavior of the free flow phase
as a repulsive-core gas.  Section IV contains our analysis of the
phase transition, long range correlations and finite-size effects.
Section V discusses how the value of $\Vmax$ affects the fluctuations
of jammed regions and how that affects the finite-size effects we see.
Our conclusions are summarized in Section VI.

\section{Methodology\label{method}}

All the simulations in this paper are done for a single lane track
with periodic boundary conditions.  The track lengths varied from
5,000 to 100,000.  We initially distributed the cars uniformly around
the track.   The system was then evolved for at least $10^6$ time steps
to form a random steady state, a time step being one update of all $N$
vehicle positions and velocities.  We then sampled the system every ten
time steps for the next $10^7$ to $10^8$ time steps, the exact length
depending on the system size.

Since this is a non-equilibrium problem, we were careful to look
for non-ergodic effects and sensitivity to initial conditions.  We used
different random seeds to generate 5-10 different steady states for each
choice of density and track length. We also did simulations using two
different random number generators.  We have seen no evidence that the
choice of initial condition or random generator affected our results,
although we have seen the need for long simulation times (much longer than
typically used) to ensure that we are seeing the steady state behavior.
If you use the ending configuration of a system at a higher density,
and use its ending configuration (minus a few cars) as a starting
configuration at lower density, you see the same results as starting
from an initially uniform distribution of cars for the lower density.
The values we show in this paper represent averages over simulation time,
initial condition and random number generator.

To analyze this model, we chose to study density correlations using the
static structure factor $S(q)$
\[
S(q) =  \langle |\rho(q)|^2\rangle \qquad
\rho(q) = \sum_{r=1}^{L} e^{-i q r} n(r)\,,
\]
and the pair correlation $G(r)$
\[
G(r) =\frac{1}{L} \sum_{q} e^{i q r} \left(\frac{S(q)}{N}-1\right)
= \left\langle \frac{1}{N}\sum_{l=1}^{L} n(l)n(l+r)\right\rangle ,
\]
where $n(r)=1$ if there is a car at site $r$ and zero otherwise.  The
angle brackets denote an average over configurations.  The other
function we examine is the nearest neighbor distribution $P(r)$
\[
P(r) =
\left\langle
\frac{1}{N}\sum_{n=0}^{N-1} \delta(r_{{n+1}}-r_{n},r)
\right\rangle
\,,
\label{nndist}
\]
where $r_{n}$ denotes the position of the n-th car and $\delta(,)$
denotes a Kronecker delta.  $P(r)$ is simply the probability that the
distance to the next car ahead is equal to $r$~\footnote{The gap between
the cars is $r-1$.}.

\section{Free Flow Regime\label{Dilute}}

In this model, the only interaction between the vehicles is the gap rule,
which comes into play only when the distance to the next car is less than
or equal to $\Vmax$.  At low density $d=N/L$, when the vehicle spacing is
typically much larger than $\Vmax$, naively applying the other dynamical
rules produces a steady state with each vehicle having a speed of $\Vmax$
or $\Vmax-1$ with a mean speed of $\Vmax-p$.  If the vehicles have this
speed distribution, the vehicle spacing evolves as a random walk with a
diffusion constant of $p(1-p)$.  However, this produces a steady state
where all spacings between cars are equally likely, including spacings
of less than $\Vmax$.  

\begin{figure}[!ht]
\includegraphics[width=0.95\hsize]{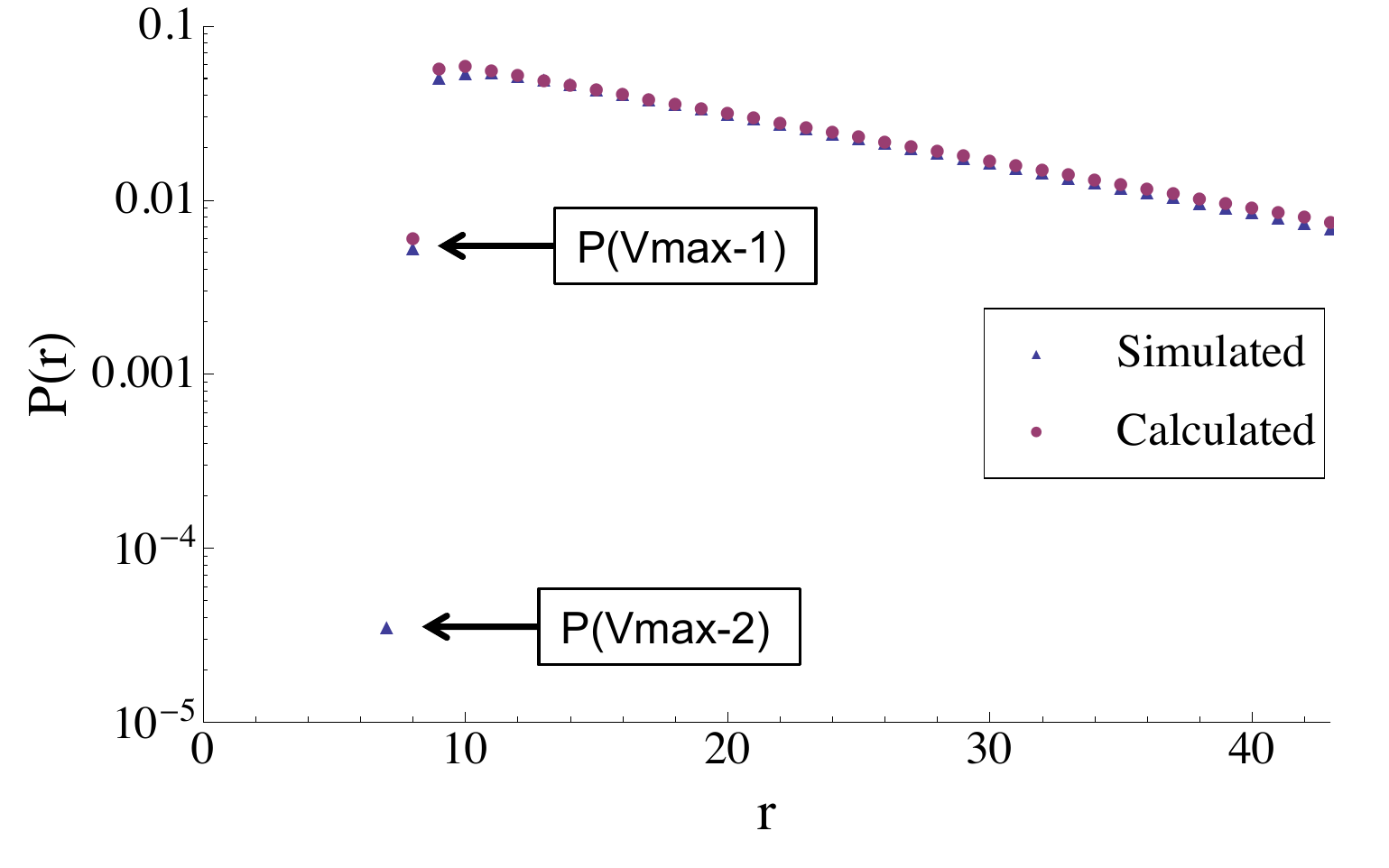}
\caption{(Color online) Semilog plot of simulated $P(r)$ and that
calculated from the Appendix for $\Vmax=9$ and $p=0.1$ at a density
$d=0.04$, about half the critical density for the jams to form.
\label{lowdensgap}}
\end{figure}

Therefore, even in the dilute regime, the gap
rule followed by the random slowdown, forces some cars to spend a small
fraction of the time at a speed of $\Vmax-2$ because the gap to the
vehicle ahead of it is $\Vmax-1$.  This vehicle will, on the next time
step, have a gap of $\Vmax-1$ or larger.  Thus each car has a ``repulsive
core'' that strongly favors at least $\Vmax-2$ empty sites ahead of it.
In Fig.~\ref{lowdensgap} we show a typical example.

If we assume that no cars have a gap of less than $\Vmax-1$, we show
in the Appendix that we can find $P(r)$ from a simple kinetic equation.
For $r>\Vmax+2$, $P(r)$ obeys a drift-diffusion Fokker-Planck equation
in the continuum limit
\begin{equation}
\frac{\partial P(r)}{\partial t}=\alpha \frac{\partial P(r)}{\partial r}
+(p(1-p)+\alpha/2)\frac{\partial^2 P(r)}{\partial r^2}
\,,
\label{dif}
\end{equation}
where $\alpha=P(\Vmax)$.  The term $\alpha/2$ is the
leading repulsive core correction to the diffusion constant.
The steady state solution to the Eq.~(\ref{dif}) for $r>\Vmax+2$ is
\begin{equation}
P(r)=P_{0} \exp\left(-\frac{\alpha}{p(1-p)+\alpha/2}\, r\right)
\,,
\label{ss}
\end{equation}
\noindent where $P_{0}$ is a constant determined from solving the
equations for $P(\Vmax+1)$ and $P(\Vmax+2)$, together with
the normalization condition $\sum_r P(r)=1$.  

There is a simple interpretation of the form of Eq.~(\ref{ss}).  Each
vehicle has an excluded region of size $\approx \Vmax$ ahead of it. If the
typical vehicle spacing is $L/N=1/d$, the effective free space between
vehicles is $1/d - \Vmax \approx p(1\!-\!p)/\alpha+ 1/2$.
An example of the agreement between the simulations and this analytic
model are shown in Fig.~\ref{lowdensgap}.

The model above assumes that no vehicles have a gap of less than
$\Vmax-1$.  The event that first results in a gap of $\Vmax-2$ requires a
configuration of {\em three} cars, each separated by a gap of $\Vmax-1$,
with the middle car then slowing down by the randomization rule while
the last car does not.  Thus we need three body interactions to see
violations of this analytic model.

Since three-body interactions are neglected, we expect that the pair
correlation function $G(r)$ in the dilute limit can be found from the
nearest neighbor distribution $P(r)$ via an Ornstein-Zernicke relation
\begin{equation}
G(r)=P(r)+\sum_{i=1}^{r-1} P(i)\, G(r-i)\,.
\label{OZ}
\end{equation}

\begin{figure}[!ht]
\subfigure[]{\includegraphics[width=0.48\hsize]{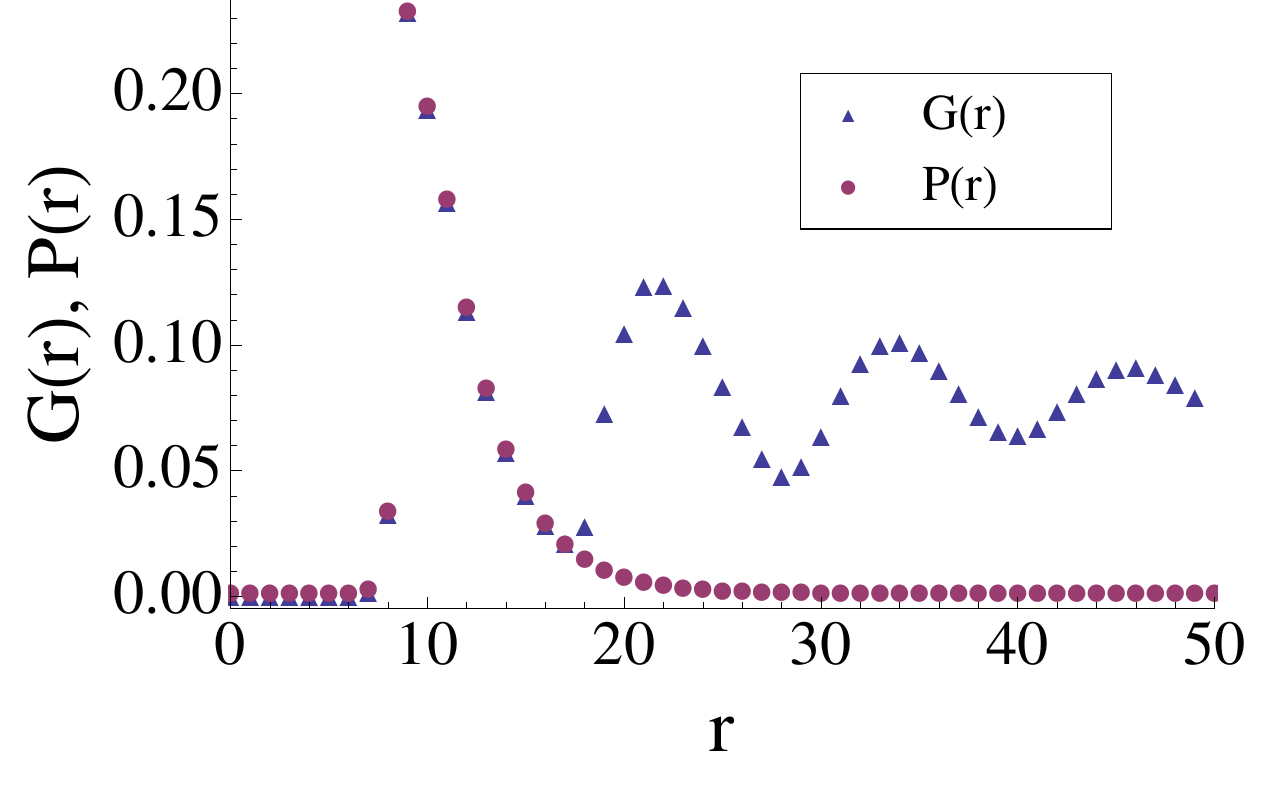} }
\subfigure[]{\includegraphics[width=0.48\hsize]{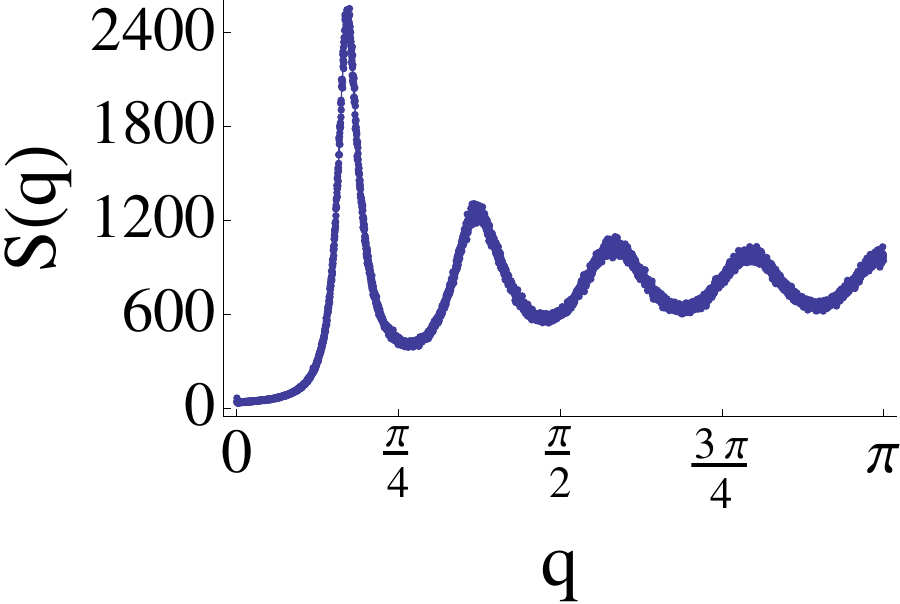} }

\caption{(Color online) (a) Simulated nearest neighbor distribution
$P(r)$ and pair correlation $G(r)$ for $\Vmax=9$ and $p=0.1$ at a
density $d=0.08$.  (b) Structure factor $S(q)$ obtained from $G(r)$
through Eq.~(\protect{\ref{OZ}}) and from simulations.
\label{DDC-Gr}}
\end{figure}

Figure~\ref{DDC-Gr}(a) shows the nearest neighbor correlation $P(r)$ and
the $G(r)$ we get from the simulations in this regime.  Since $P(r)$ is
vanishingly small for $r<\Vmax$, Eq.~(\ref{OZ}) predicts that $P(r)$ and
$G(r)$ are identical up to $r=2\Vmax$ which Fig.~\ref{DDC-Gr}(a) shows.
The $G(r)$ that we find from the Eq.~(\ref{OZ}) is indistinguishable from
the simulations.  Figure~\ref{DDC-Gr}(b) shows the corresponding structure
factor $S(q)$.  The peaks in $S(q)$ at multiples of $q=2\pi/\Vmax$
are simply the result of the repulsive core seen in $G(r)$.  We note
for future reference that $S(q)$ shows no upturn at $q\to0$, indicating
there is no long range order in the dilute regime.

\section{Finite Size Effects: Long-Range Correlation\label{Finite}}

\begin{figure}[!ht]

\subfigure[]{\includegraphics[width=0.48\hsize]{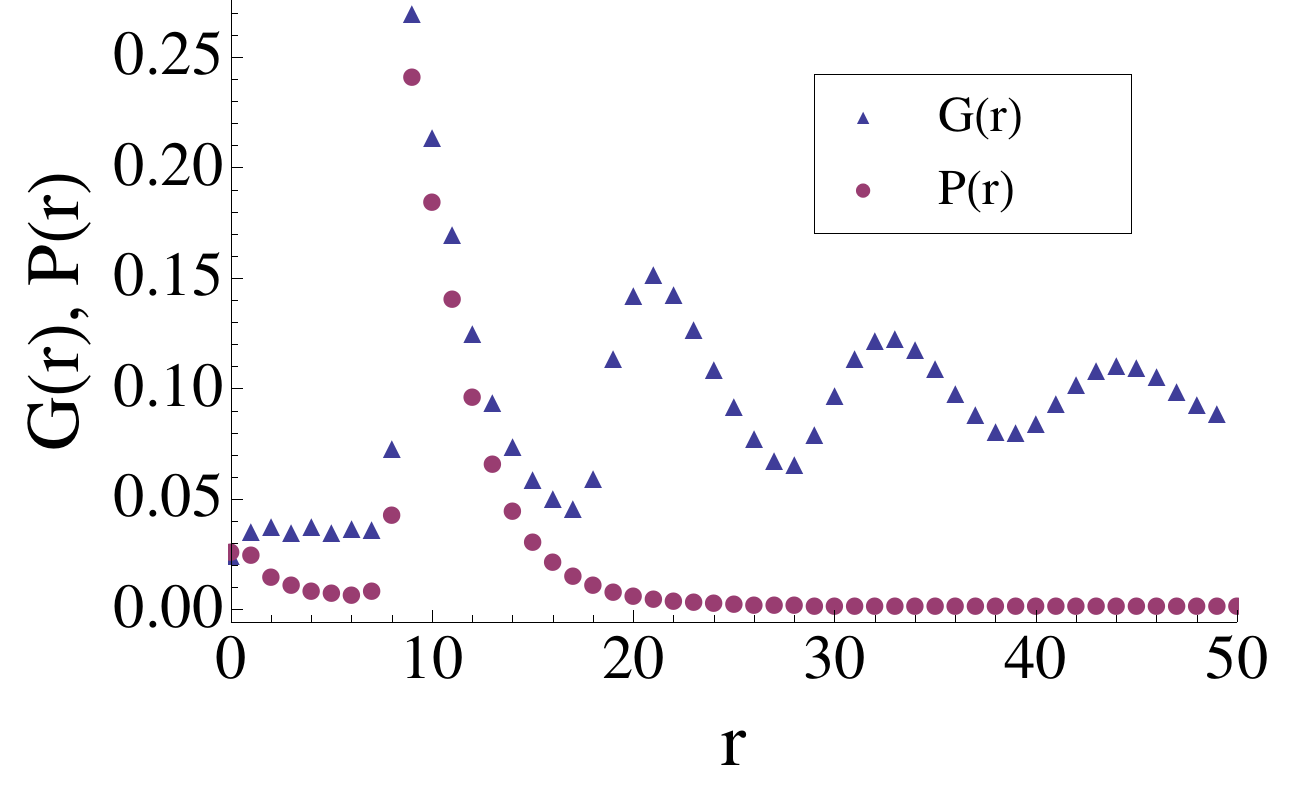} }
\subfigure[]{\includegraphics[width=0.48\hsize]{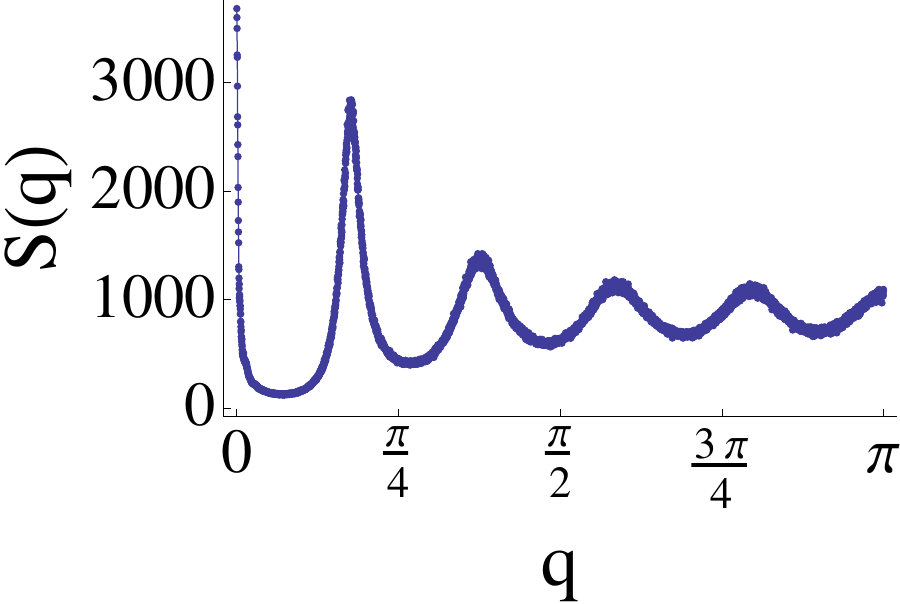} }

\caption{(Color online) (a) Nearest neighbor distribution $P(r)$ and pair
correlation $G(r)$ for $\Vmax =9$ and $p=0.1$ for a density $d=0.088$,
just above the density where jams appear.  (b) Corresponding structure
factor $S(q)$ with an upturn near $q\to 0$.\label{SF-Sq}}

\end{figure}

As the vehicles density is raised, the repulsive core gas description 
we developed above
remains qualitatively correct, with a gradual growth in the number of
vehicles spaced at shorter distances $\Vmax-2$, $\Vmax-3$, \dots.
When the jams appear, we see an abrupt change in the shape of the
nearest neighbor distribution $P(r)$ with the sudden appearance of a
nonzero fraction of vehicles with $r=1,2$.  As Fig.~\ref{SF-Sq}(a) shows,
the pair correlation $G(r)$ no longer agrees with $P(r)$
for $r\leq 2\Vmax$, and the Ornstein-Zernicke relation (\ref{OZ}) between
the two no longer holds.  At the same time, Fig.~\ref{SF-Sq}(b) exhibits an
upturn in $S(q)$ for $q\to 0$, indicating the appearance of long range
correlations in the density.

We interpret this as indicating that the free flow phase is still stable,
but that we have nucleated a new phase of localized jams that appear and
disappear.  Indeed, by examining the permanent stability of a localized
jam, Gerwinski and Krug~\cite{Gerwinski1999} have shown that the jams
should be stable at a density $\ge (1-p)/(1+\Vmax -2p)$,
which is a higher density than where we see the onset of jams.

\begin{figure}
\includegraphics[width=0.95\hsize]{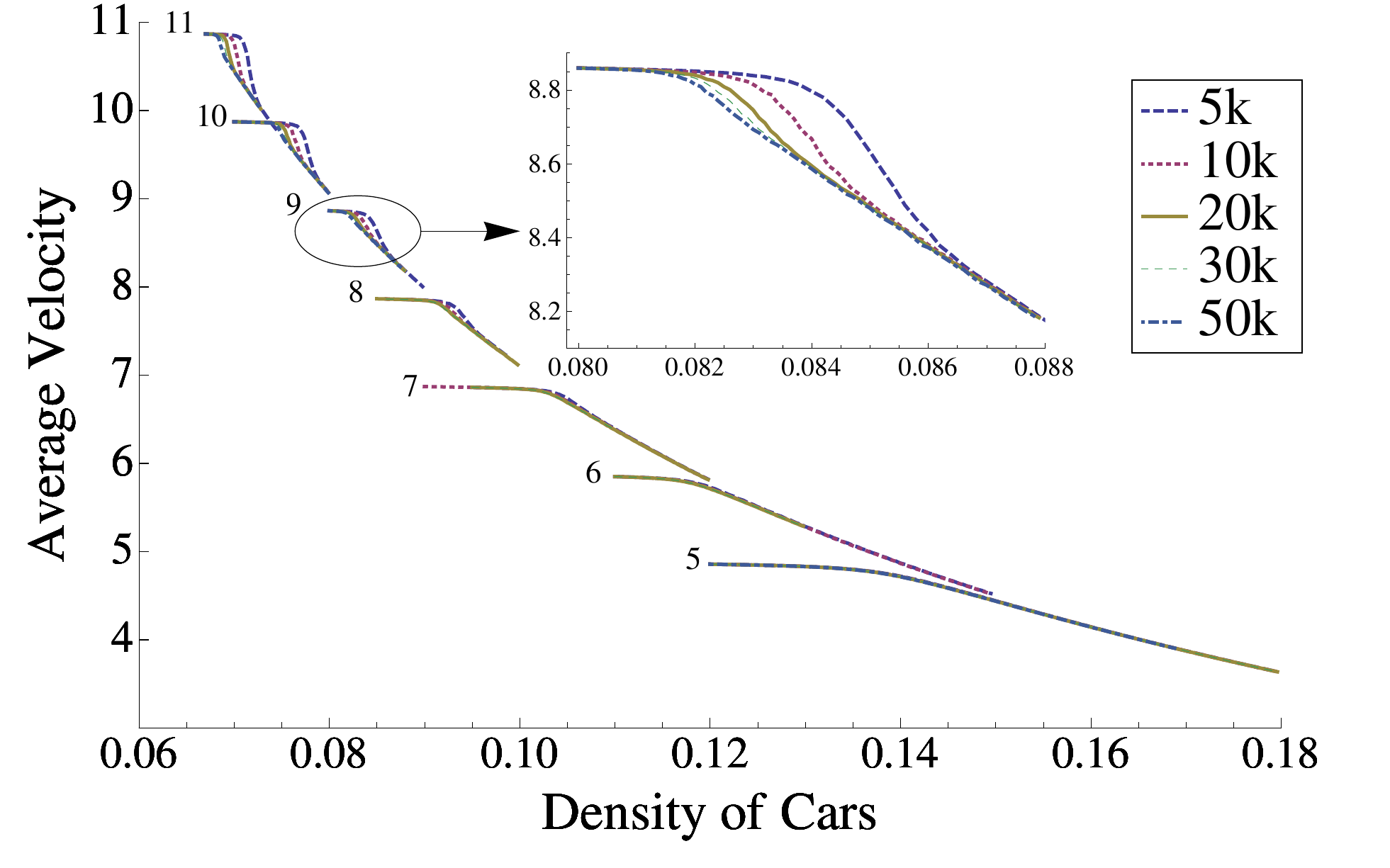}

\caption{(Color online) The dependence of the mean velocity on density
for various values of $\Vmax$ and $p=0.1$, and a variety of track lengths
(shown as different colors).\label{globalvel}}

\end{figure}

The upturn in $S(q)$ for small $q$ indicates some long-range order,
which implies that we might observe finite-size effects in various
quantities that are sensitive to the presence of that long range order.
Figure~\ref{globalvel} shows how the average velocity changes with
density for different values of $\Vmax $ and different track lengths.
For $\Vmax\alt 6$ we observe no length dependence.  The figure also
shows that once the density is well above the transition density, the
system is insensitive to both the value of $\Vmax$ and the system size.

\begin{figure}
\includegraphics[width=0.95\hsize]{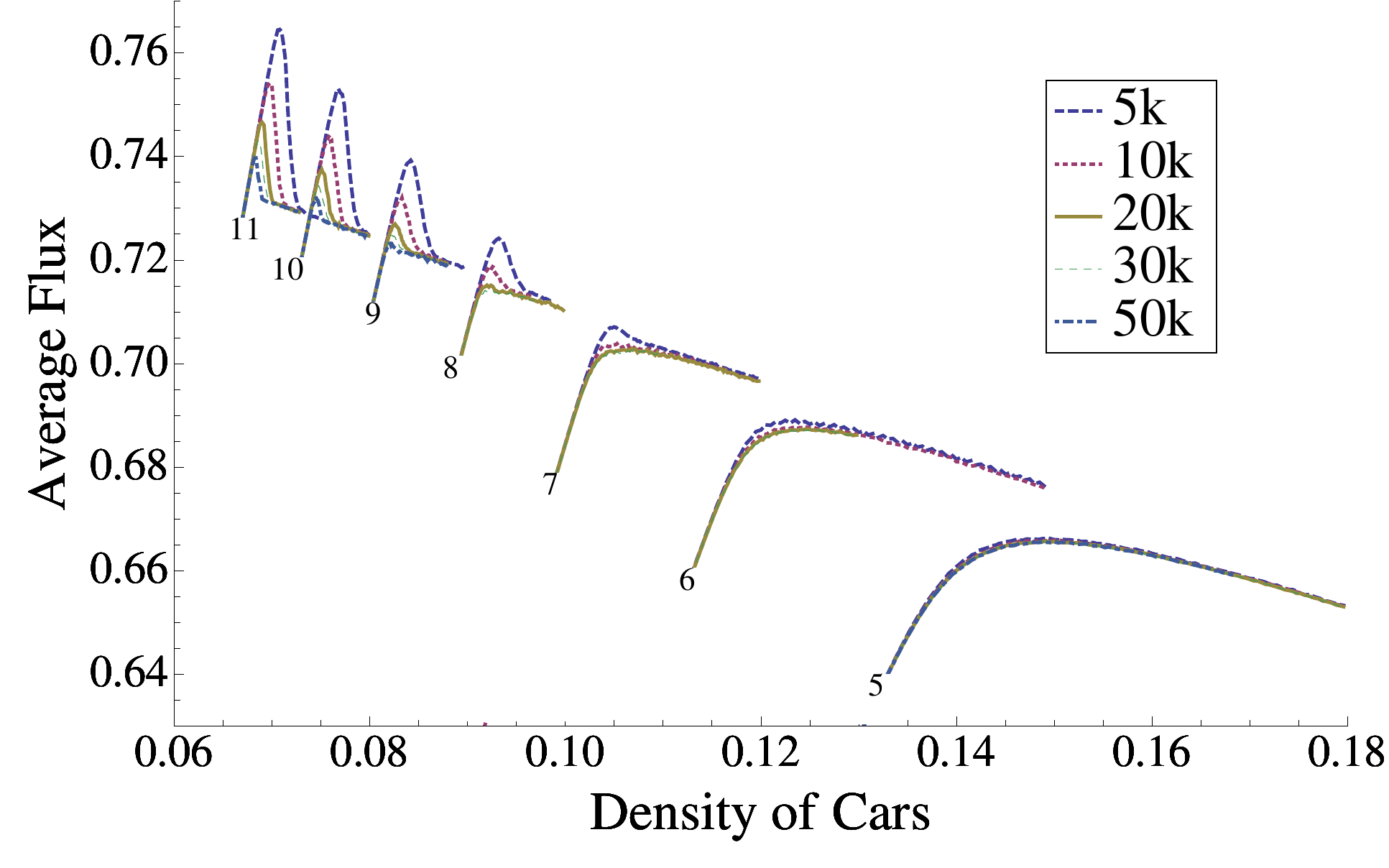}

\caption{(Color online) The average flux versus density for various
$\Vmax$ with $p=0.1$ and track lengths (different colors) of $L=$ 5k,
10k, 20k, 30k, 40k, 50k.
\label{flux}}

\end{figure}

This size sensitivity is even more apparent in the mean flux of vehicles,
presented in Fig.~\ref{flux}.  As in the previous figure, it is only for
$\Vmax\agt 6$ that we see this size sensitivity.  It also means that in a
system of smaller size, the vehicle flux is actually higher than it is
in larger systems, and that the size of the effect depends on $\Vmax$.
This behavior is the reverse of what one would expect from 
hysteresis, where a large system would get trapped in a high flux 
free flow regime while a smaller system would not.

Since the value of $\Vmax$ represents the number of degrees of freedom for
each car, it is not surprising that the finite-size behavior can depend
on the number of degrees of freedom, as it does in equilibrium systems.
However, we do not have any clear evidence that there is a critical
value of $\Vmax$ for which the finite-size effects appear, but they are
clearly suppressed for $\Vmax\leq 5$.

To characterize the transition, we need a quantity sensitive to the
presence of jams.  We discussed in the Introduction a variety of choices
that others have used that are based
on the velocity distribution.  In this study, where we focus on the
spatial distribution of the cars, we have used the gaps rather than
the vehicle speeds to characterize the jams.  The gap rule, however,
produces a strong correlation between the speed of a vehicle and the
distance to the next car, so our order parameter is closely related
to these other choices.

\begin{figure}
\includegraphics[width=0.95\hsize]{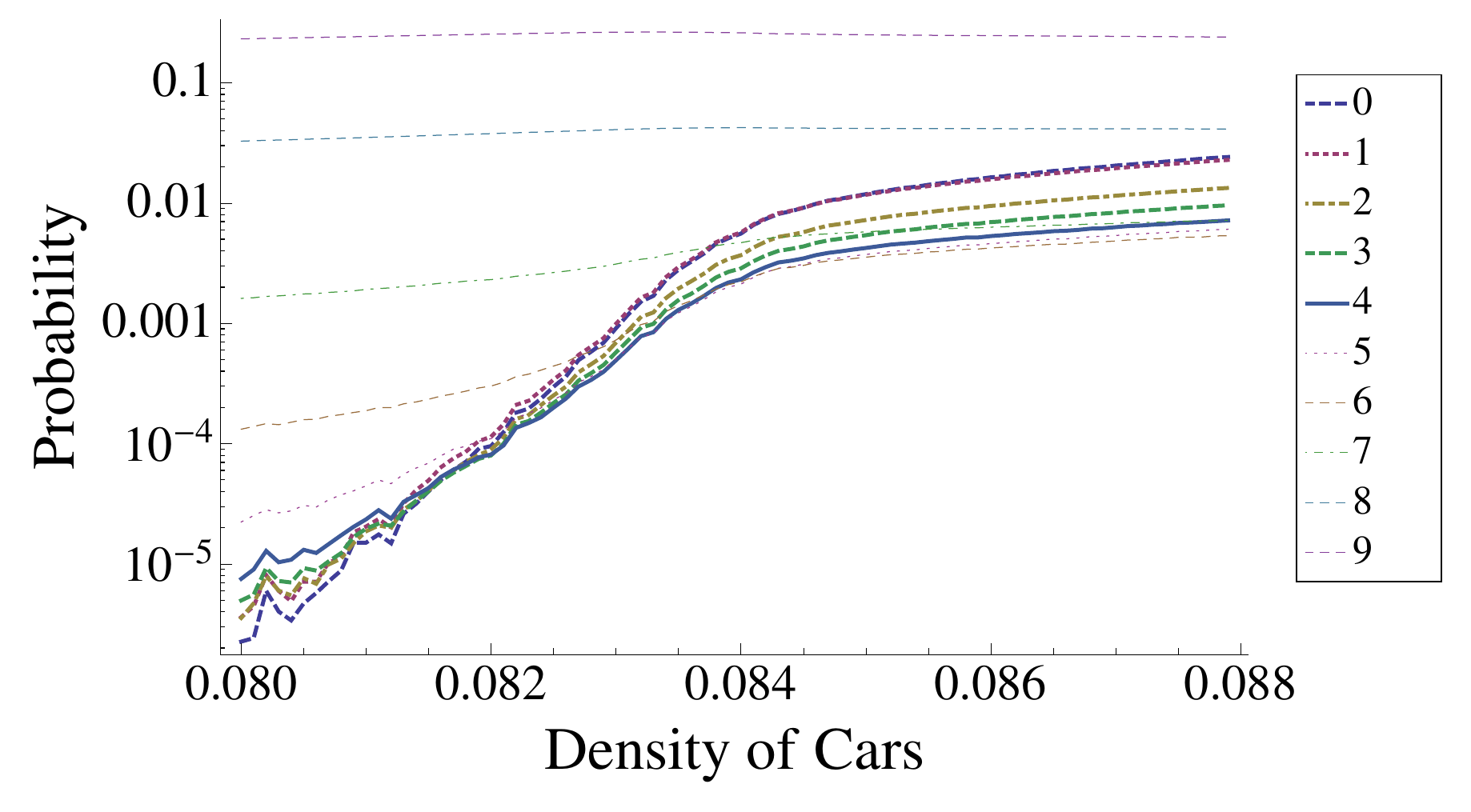}
\caption{(Color online) 
Probability of finding a gap of a particular size for various
densities for $\Vmax=9$ and $p=0.1$.  \label{allgap}}
\end{figure}
Figure~\ref{allgap} shows how the probability of finding gaps of
different sizes varies with density near the transition.  We see that
the probability of having a gap $\leq \Vmax/2$ changes dramatically here.
Therefore, we will define the order parameter $x_{0}$ to be the fraction
of vehicles with a gap $\leq \Vmax/2$.  We could have used just the
vehicles with a gap of zero~\cite{Vilar1994}, but using all
of these gaps gives us more reliable statistics.

\begin{figure}
\subfigure[]{\includegraphics[width=0.95\hsize]{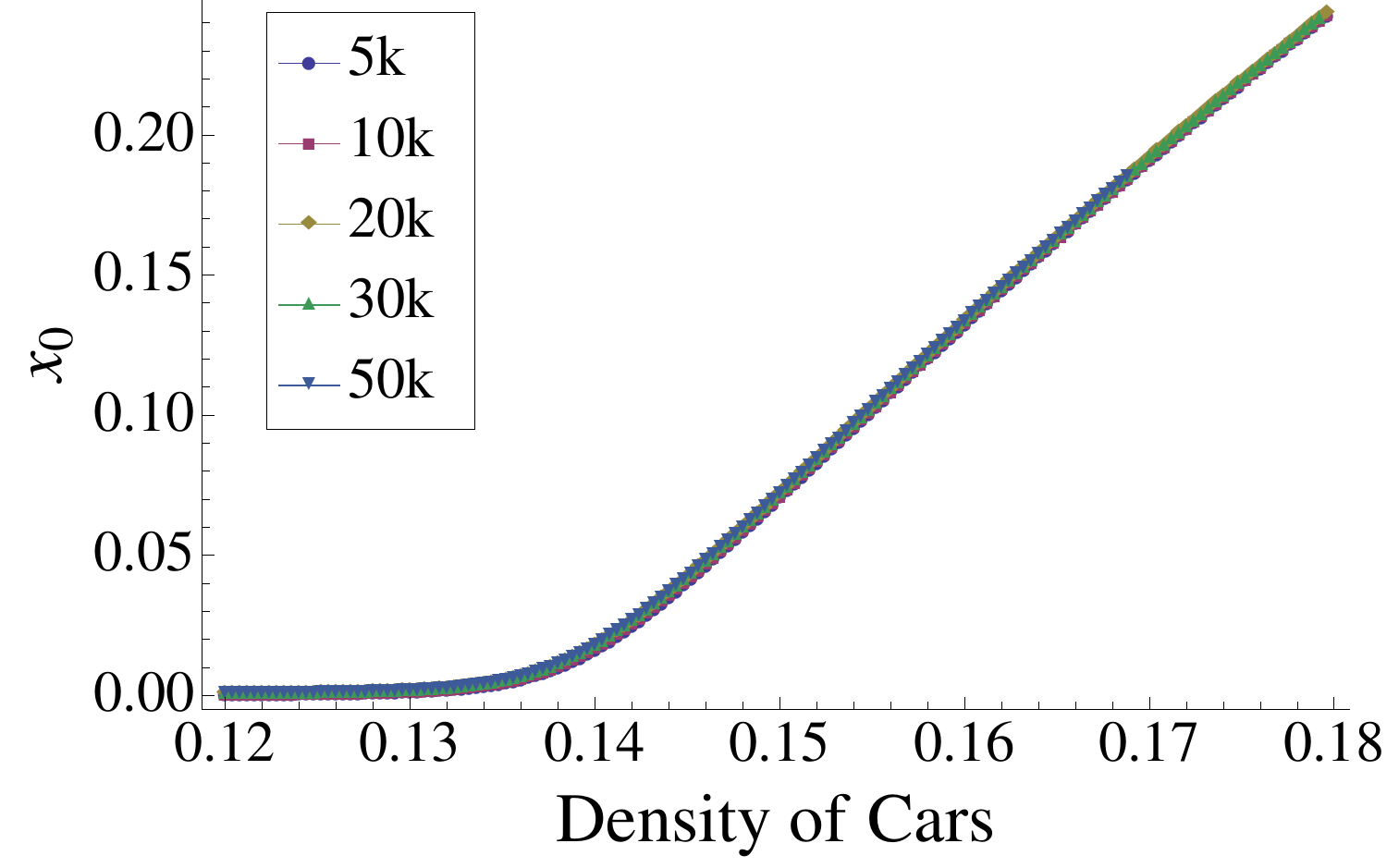} }
\subfigure[]{\includegraphics[width=0.95\hsize]{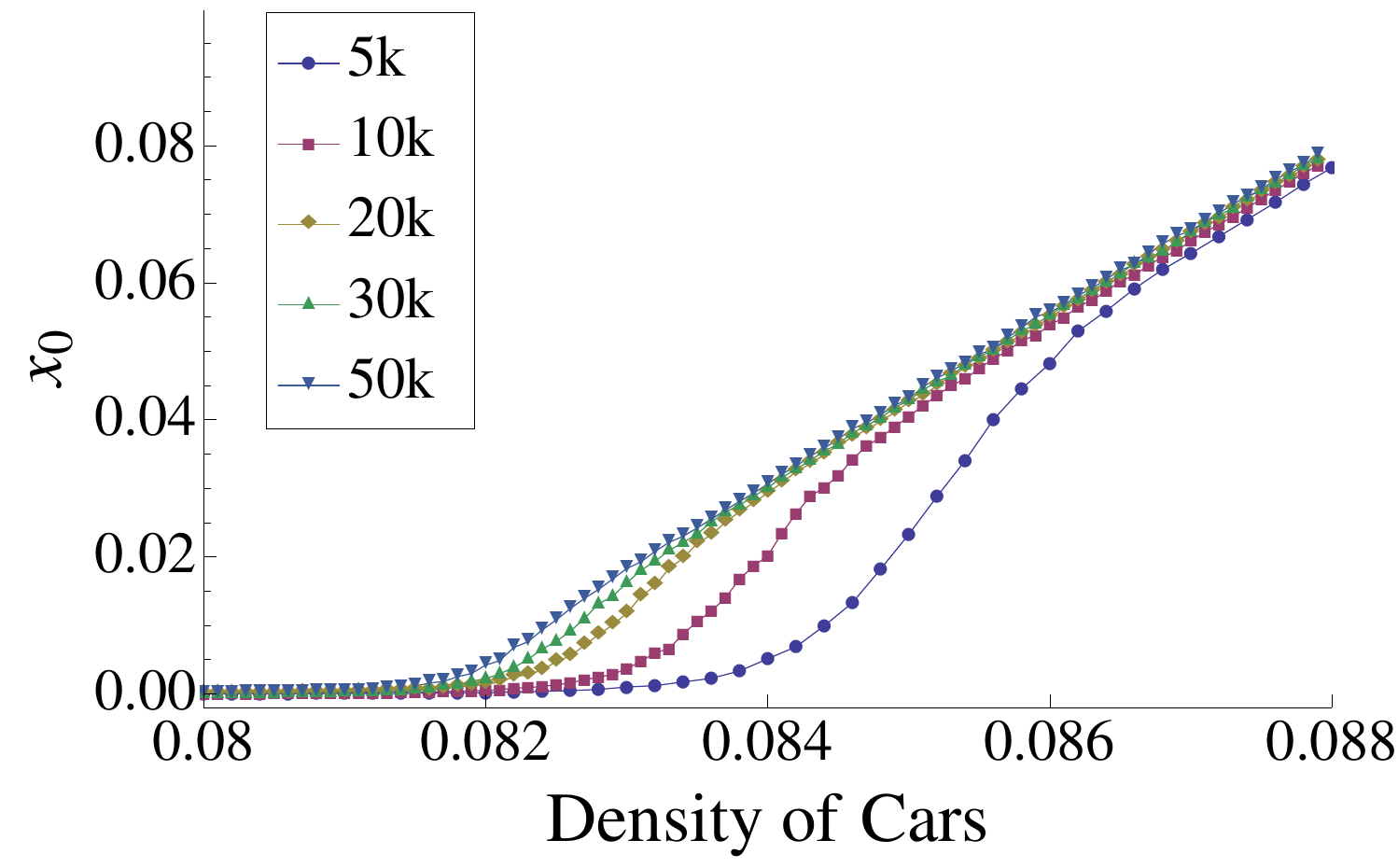} }
\caption{(Color online) 
Plots of the probability of having gap $\leq\Vmax /2$ for different
track lengths and $p=0.1$ for (a) $\Vmax =5$ and (b) $\Vmax =9$.
\label{gapsize}}
\end{figure}
Figure~\ref{gapsize} shows that this order parameter $x_{0}$ exhibits
the same finite-size effects, including its dependence on $\Vmax$,
that we observed for the mean flux and velocity.  These finite-size
effects observed for $\Vmax\agt 6$ are most pronounced for small systems,
not large systems, unlike the hysteresis seen in an equilibrium first
order transition.

In order to examine the long range correlations in this transition,
we use a finite-size scaling approach~\cite{Fisher1967,Barber1983,
Ferrenberg1991}.  Accordingly, we assume that the order parameter $x_{0}$
near the transition point depends on the size of the system as:
\begin{equation}
x_{0} = L^{-a_{0}} f(L/\xi)\,,
\label{scalingeq}
\end{equation}
\noindent where $a_0$ is the scaling exponent, and
$\xi$ is the correlation length, which itself depends on $L$ via
\begin{equation}
\xi \propto (d- d_{c}(L))^{-\nu}\,,
\label{xi}
\end{equation}
\noindent where $ d_{c}(L)$ is the critical density.

In most finite-size scaling studies, the transition point itself is
dependent on the system size.   That effect is usually considered
as a correction to scaling~\cite{Ferrenberg1991}, and $d_{c}(L)$ then
considered independent of $L$.  We found that we got much better scaling
fits by considering a length dependent critical density via
\begin{equation}
 d_{c}(L)=d_{0} + c L^{-b}\,,
\label{dc}
\end{equation}
where $d_{0}$ is the critical density for the infinite system.  We have
tested~\cite{Thesis} this approach on the 3d Ising model, which was
studied by Ferrenberg and Landau~\cite{Ferrenberg1991} in their study
of finite-size scaling corrections, and we find excellent agreement with
their results.

To find the shift in the critical density, the typical
approach~\cite{Ferrenberg1991} is to study a quantity like a
susceptibility that has a peak at the transition.  Our order
parameter has no maximum at the transition, so we study its
derivative~\cite{Ferrenberg1991} instead.  While at first thought, a
quantity like $\langle x_{0}^2\rangle$ might act like a susceptibility,
we have found that this jamming transition is not like an equilibrium
transition with large fluctuations in the order parameter correlations
before the transition.  Instead, we are seeing the nucleation of a
different phase (the jams) in a background of the free phase, and the
fluctuations in $x_{0}$ basically track $x_{0}$.

Instead, we examined the dynamic susceptibility, $\chi_{4}$, which
was used to study glassy behavior in the NS model in the $p\to1$
limit~\cite{DeWijn2012}  
\begin{equation}
\chi_{4} =\frac{1}{\langle{v^2}\rangle-\langle{v}\rangle^2}\,
\left\langle \frac{1}{N}\sum_{i=1}^{N}
\sum_{j=0}^{N}( v_{i}-\overline{v})(v_{j}-\overline{v})\right\rangle \,,
\label{param}
\end{equation}
where $v_{i}$ is the velocity of the i-th car at a particular
time and $\overline{v}$ denotes the mean speed of all the cars at
that time.  $\chi_{4}$ measures the number of vehicles that move
cooperatively~\footnote{The expression for $\chi_{4}$ here is the equal
time correlation $\chi_{4}(0)$ used in Ref.~\cite{DeWijn2012}.}.

\begin{figure}
\includegraphics[width=0.95\hsize]{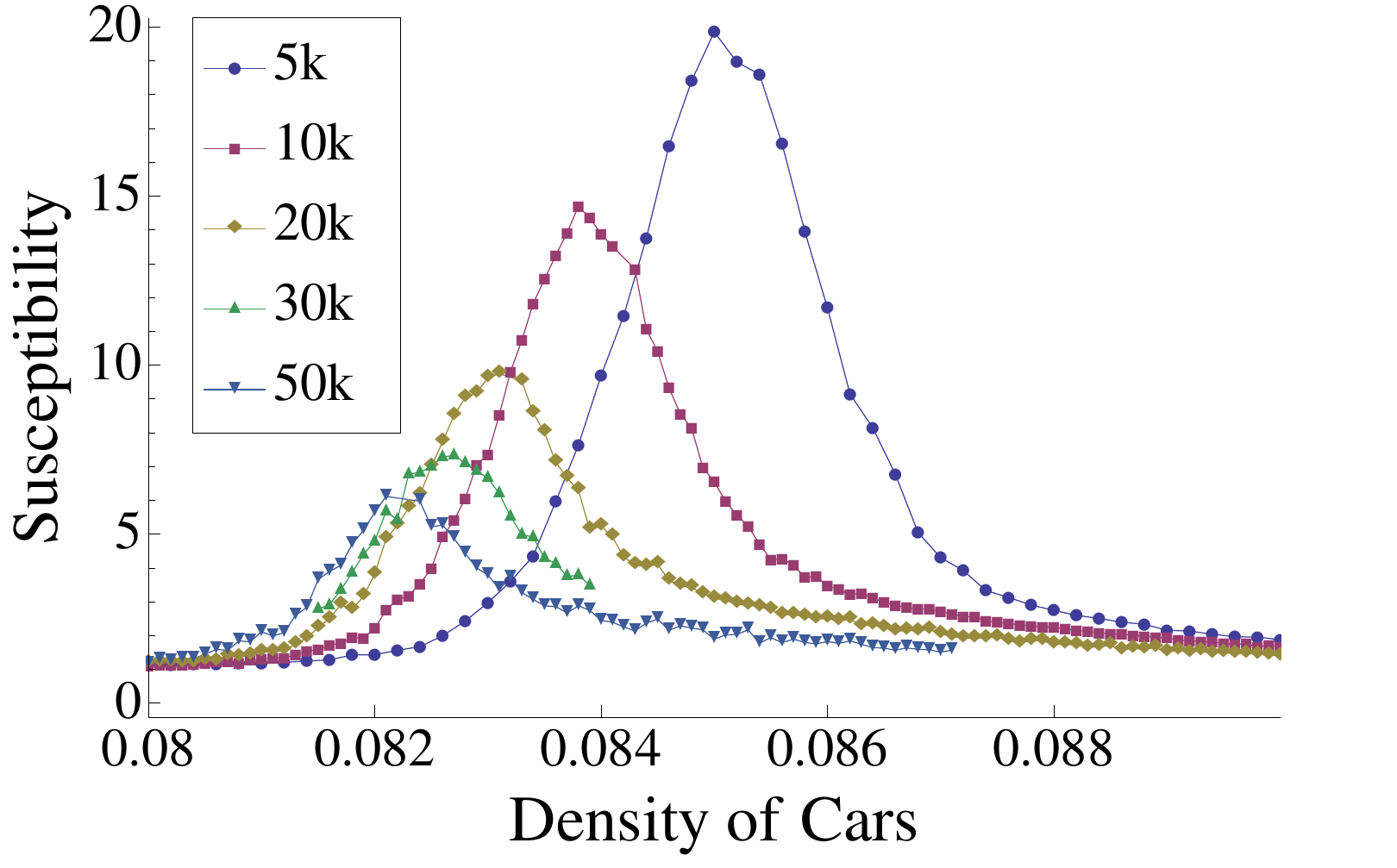}
\caption{(Color online) The dynamic susceptibility $\chi_{4}$ for $\Vmax
=9$, $p=0.1$ and various system sizes.
\label{chi4}}
\end{figure}

As Fig.~\ref{chi4} shows, $\chi_{4}$ has finite-size effects for
$\Vmax\agt 6$ with a peak as it goes through the transition.  We assume
$\chi_4$ obeys a finite-size scaling form
\[
\chi_4 = L^{-a_4}f(L/\xi)\,.
\]
Using the peak in $\chi_{4}$ in Fig.~\ref{chi4} as the critical density,
we find for $\Vmax =9$ and $p=0.1$ that the bulk transition occurs at
$d_{0}=0.08122 \pm 0.00004$, and the shift in the transition due to the
finite system size has an amplitude of $c=0.375 \pm 0.006$ and a scaling
exponent of $b=0.54 \pm 0.02$.

If we instead calculate the derivative of our order parameter $x_{0}$
and use its peak to calculate the shift of the transition point, we
find $d_{0}=0.08267 \pm 0.00001$, $c=9.415 \pm 5.278$, and $b=0.956 \pm
0.065$.  There is no reason to expect that the peak in the derivative
of $x_0$ should be at the same place as the $\chi_{4}$ peak, so it is
not surprising the two results give different results for the shift in
the transition point.

\begin{figure}
\subfigure[]{\includegraphics[width=0.95\hsize]{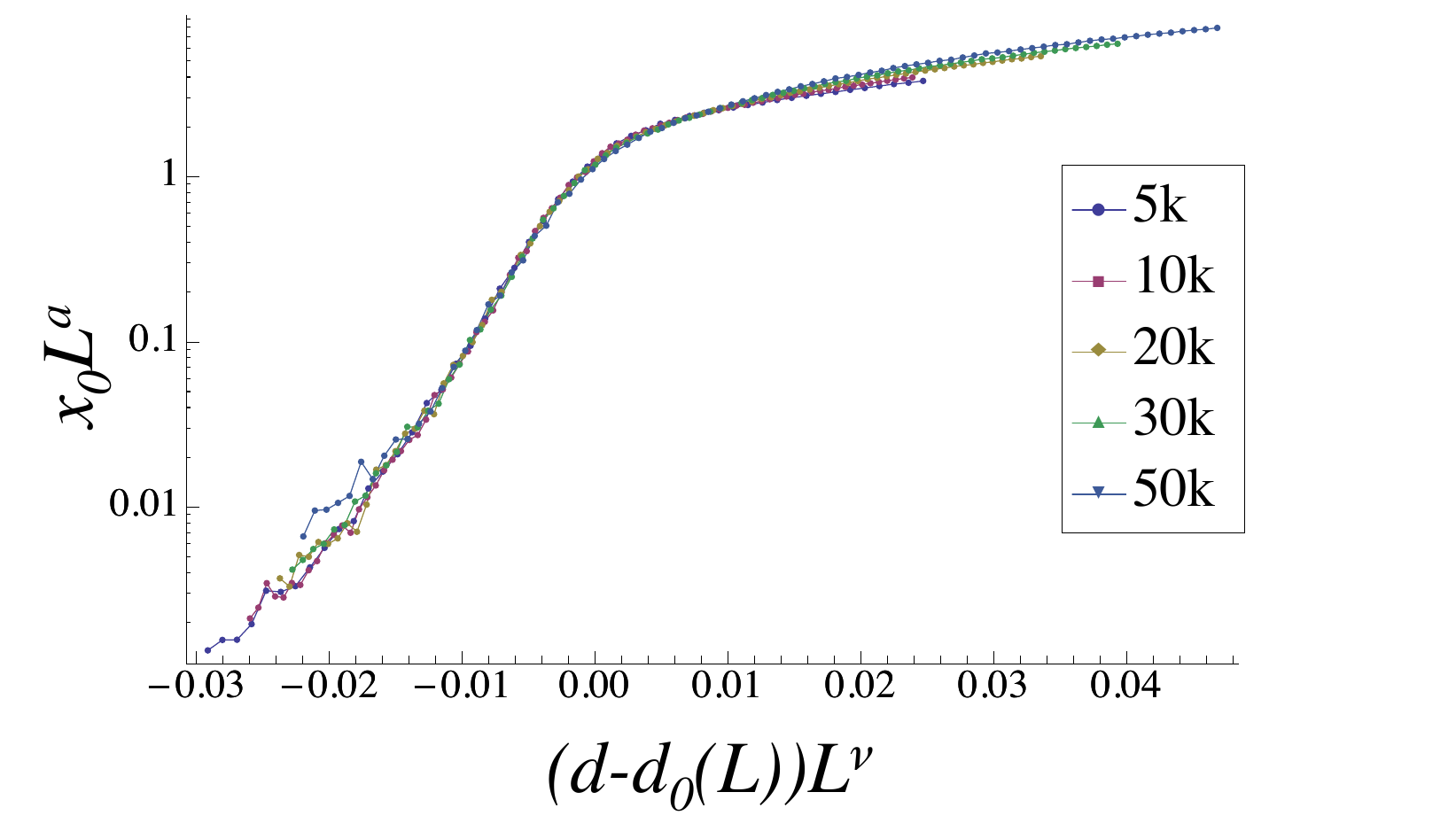} }
\subfigure[]{\includegraphics[width=0.95\hsize]{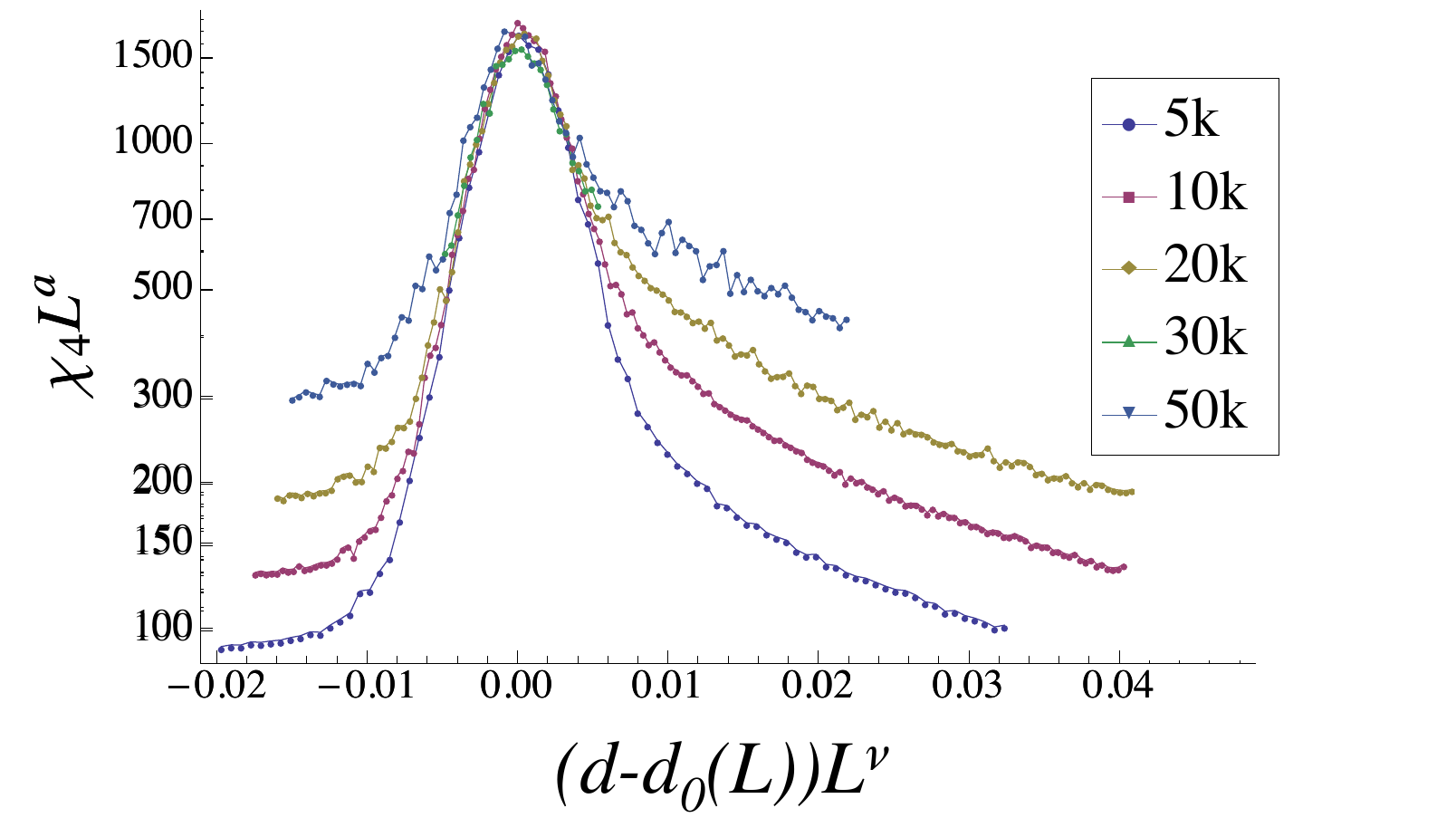} }

\caption{(Color online)
Scaling plots of (a) $x_{0}L^{a_0}$ and (b) $\chi_{4}L^{a_4}$
versus $(d-d_c(L))L^{\nu}$ for $\Vmax=9$ and $p=0.1$.
\label{finitesize}}

\end{figure}

After determining the shift in the transition, we collapse the data onto a
single curve by plotting the quantity versus $(d-d_c(L))L^\nu$ and adjust
the exponents to minimize the area bounded by the scaled data.  The scaled
plots of $x_{0}$ and $\chi_{4}$ are shown in Fig.~\ref{finitesize}.  No
matter which method we use to determine the shift in the transition point,
we find that $x_{0}$ and $\chi_{4}$ produce values for the correlation
length exponent $\nu$ of $0.13\pm0.02$ and $0.14\pm0.02$ respectively.
The scaling exponent for the amplitude of $x_{0}$ is $a_0 =0.24\pm0.04$
and that of $\chi_{4}$ is $a_4=0.52\pm0.02$.  We also examined the scaling
behavior of several alternative order parameters: the probability of a
car having a speed $<\Vmax/2$~\cite{Jost2003}, the difference between
the mean speed and that of the free flow speed~\cite{Souza2009}, and
the difference between the variance in the velocity and its value in
the free flow regime~\cite{Thesis}.  All of them gave results for $\nu$
and the scaling amplitude exponent $a_0$ that were consistent with those
found for $x_{0}$.

\begin{figure}
\includegraphics[width=0.85\hsize]{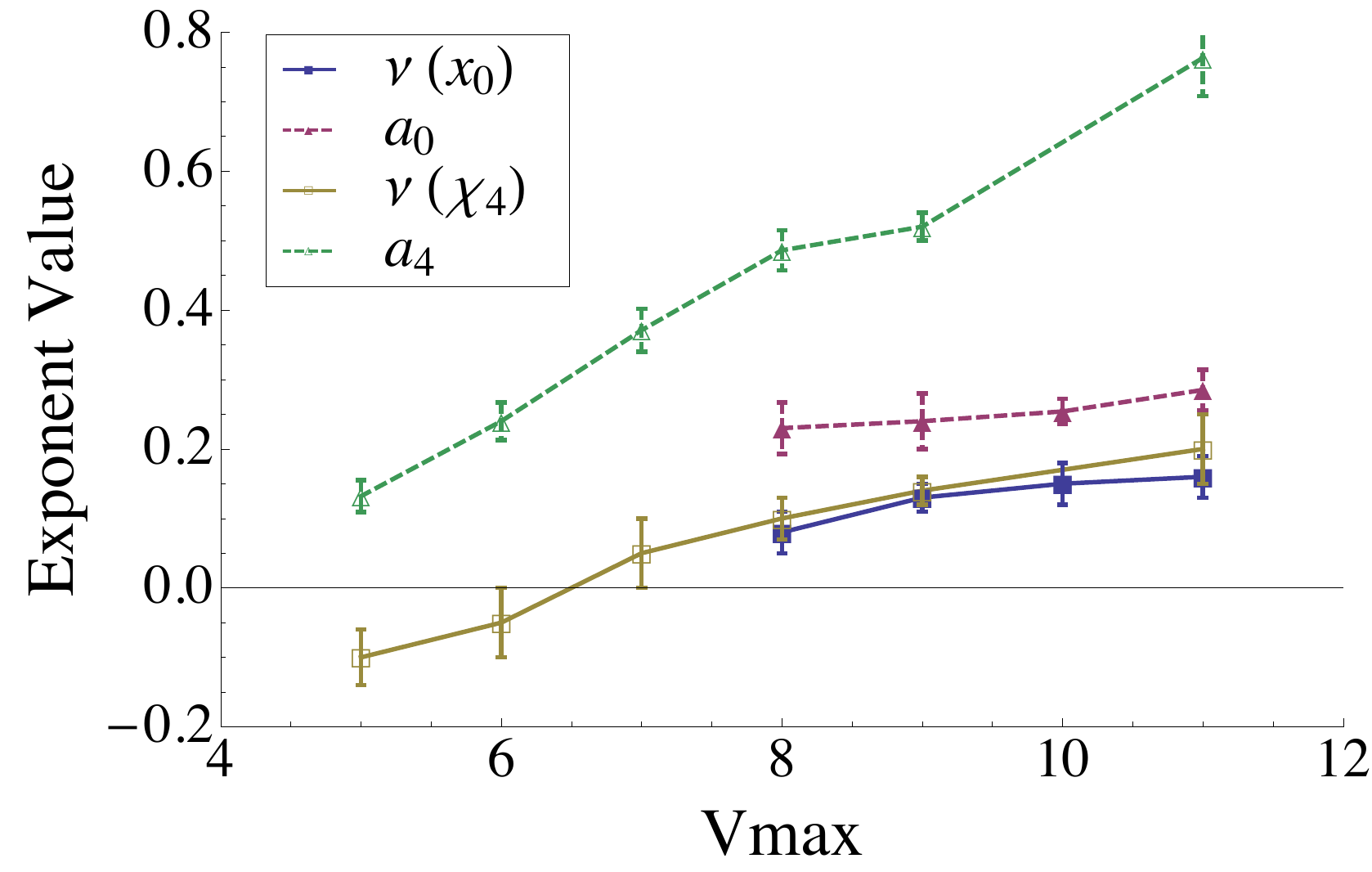}
\caption{(Color online)
Exponents $\nu$ and $a$ for various values of $\Vmax$, with $p=0.1$.
The smaller error bars on $\Vmax=9$ result from using more track sizes.
\label{exponents}}
\end{figure}

We determined the values of $\nu$ and the scaling exponents for the
amplitude of $x_{0}$ and $\chi_{4}$ for a range of values of $\Vmax$
and $p$.  The values of the exponents for different values of $\Vmax$
and $p=0.1$ are shown in Fig.~\ref{exponents}.  The scaling behavior
of $\chi_{4}$ and $x_{0}$ both yielded values for $\nu$ that were
statistically the same for $\Vmax>7$.  For $\Vmax=6$ and $\Vmax=7$,
the finite-size dependence was so weak that we could not get reliable
values for the scaling amplitude of $x_{0}$, and we were only able to
extract a value for $\nu$ from $\chi_{4}$.

While this data does not imply a sharp change in the scaling behavior for
$\Vmax=6$, it does indicate that the finite-size effects for $\Vmax>7$
are completely different than for $\Vmax\leq 6$, which is already apparent
in Figs.~\ref{globalvel} and \ref{flux}.  Figure~\ref{exponents} clearly
shows that the exponent $\nu$ appears to vanish or take on unphysical
negative values for $\Vmax=5$ and 6, indicating that the long range
correlations are absent below $\Vmax=7$.

Table~\ref{tab1} shows the scaling exponents for $\Vmax=9$ and three
values of $p$, and we see no significant variation of $\nu$ or $a_0$
with $p$.
\begin{table}[ht]
\begin{tabular}{*{3}{|c}|}
\hline
$p$  & $\nu$ & $a_{0}$ \\
\hline
 0.1  &  $0.13 \pm 0.02$ & $0.24 \pm 0.04$ \\
 0.2  &  $0.12 \pm 0.02$ & $0.26 \pm 0.04$ \\
 0.5  &  $0.14 \pm 0.02$ & $0.30 \pm 0.01$ \\
\hline
\end{tabular}
\caption{(Color online)
Exponents $\nu$ and $a_0$ for $\Vmax=9$ and various values of $p$.}
\label{tab1}
\end{table}
The results agree with our expectation that we do not observe any
variation of the exponents for $0<p<1$, since the value of $p$
controls the amount of stochastic behavior and the rate the system
evolves through its configurations.  Of course, in the special
limits $p\to0$~\cite{Zhang2011} and $p\to1$~\cite{DeWijn2012} glassy,
irreversible, behavior is observed instead.

\section{Growth of the Jams\label{Transition2}}

As we noted earlier in Fig.~\ref{SF-Sq}, the finite-size effects that
appear at the transition are accompanied by an upturn in the structure
factor $S(q)$ for small $q$.  It is therefore tempting to see whether
the upturn in $S(q)$ merely reflects long range spatial correlations
in $x_{0}$.  We therefore define a local density $n_{0}(r)$ which is 1
if a vehicle is at site $r$ and the gap to the next car ahead is less
than $\Vmax/2$.  The extensive quantity $N_{0}=\sum_r n_{0}(r) \equiv
x_{0}N$ provides a measure of the number of cars participating in a jam.

\subsection{Spatial Correlations of Jams\label{DDCJ}}

\begin{figure}
\subfigure[]{\includegraphics[width=0.95\hsize]{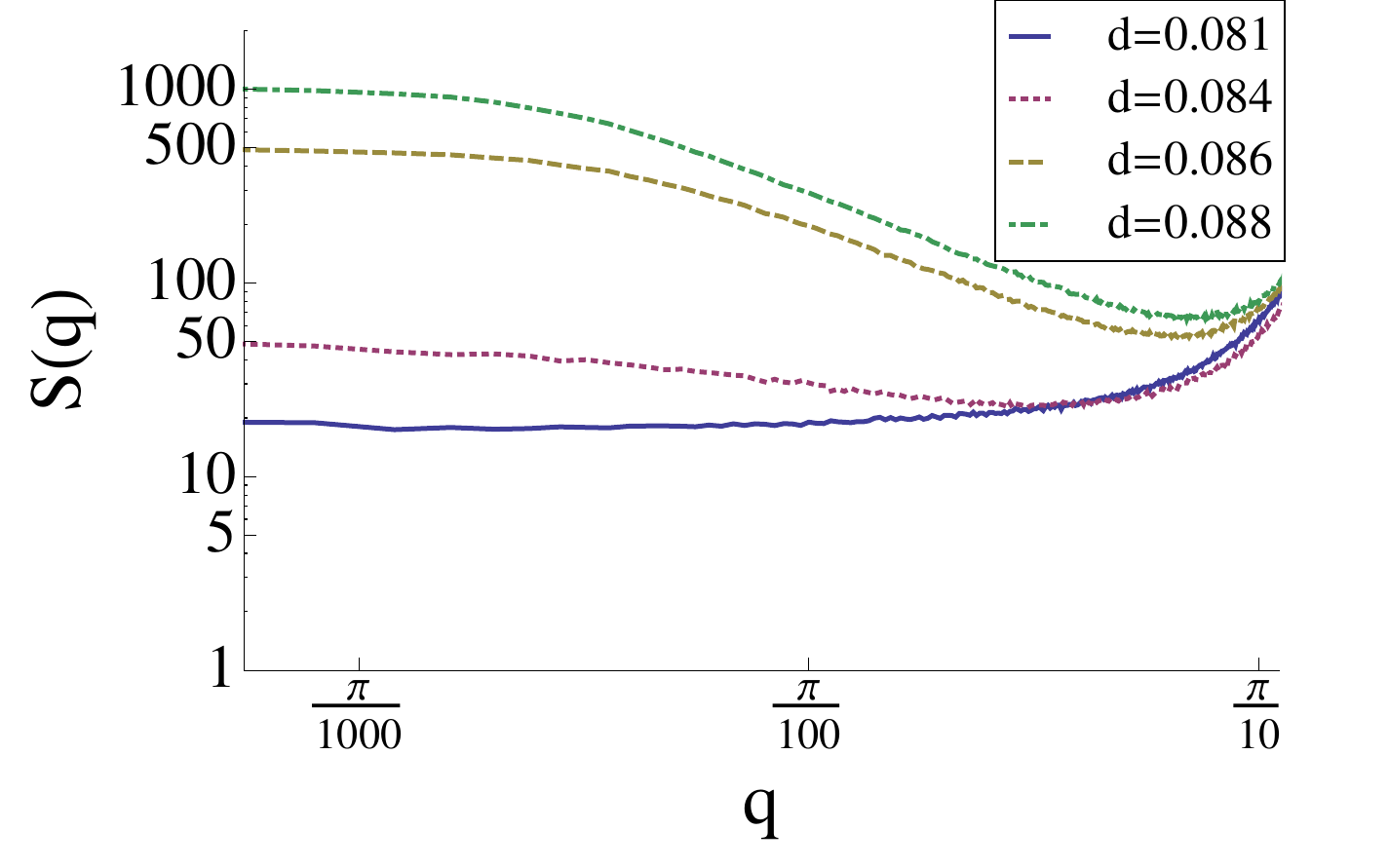} }
\subfigure[]{\includegraphics[width=0.95\hsize]{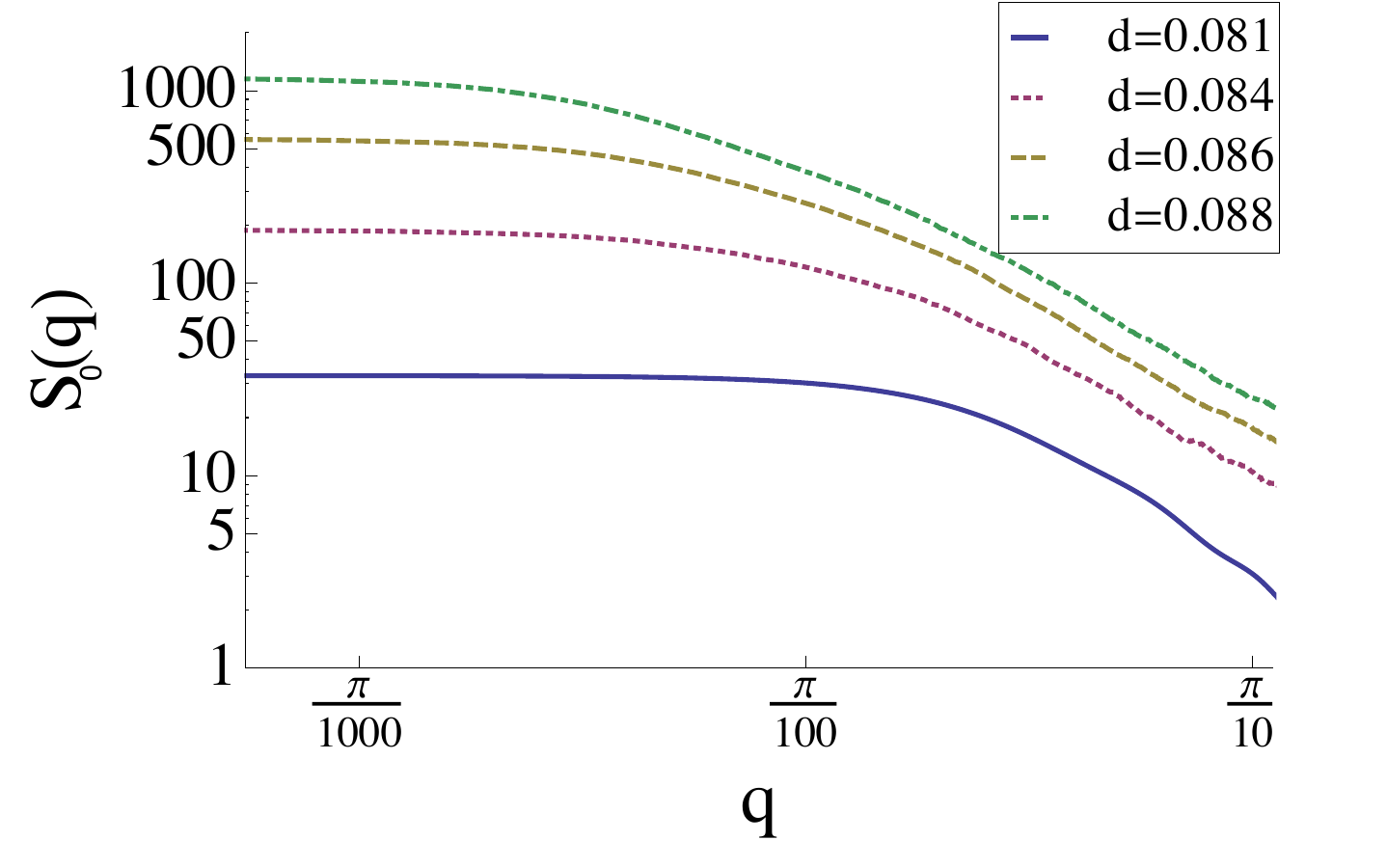} }

\caption{(Color online) 
Behavior of (a) the total structure factor $S(q)$ and (b) the structure
factor for jammed vehicles $S_{0}(q)$ for small $q$.  The data are for
$\Vmax=9$, $p=0.1$ and $L=5000$.  The densities correspond to before
the transition, in the transition region, and beyond the transition.
\label{S0upturn}
}
\end{figure}

\begin{figure}
\includegraphics[width=0.45\hsize]{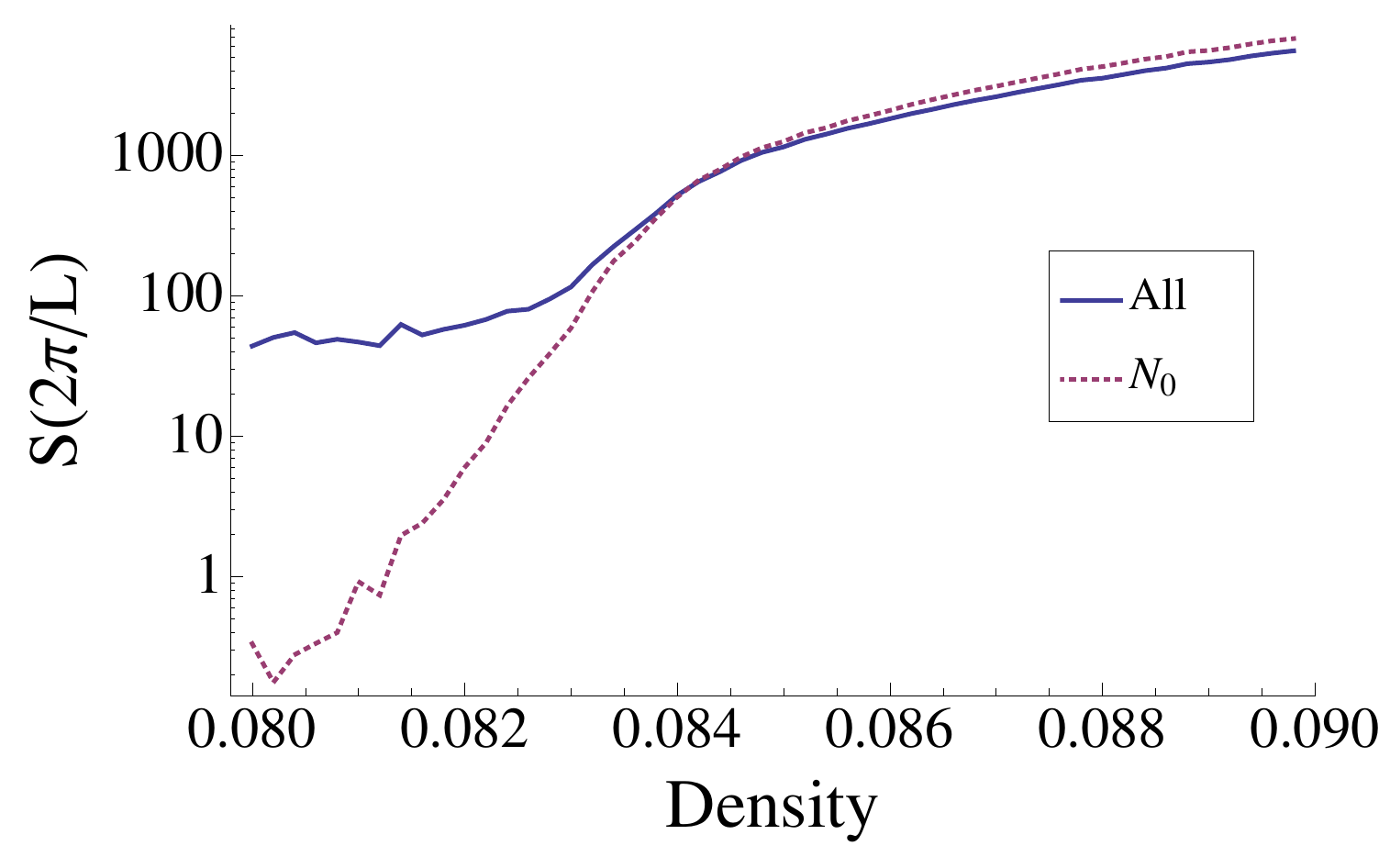}
\includegraphics[width=0.45\hsize]{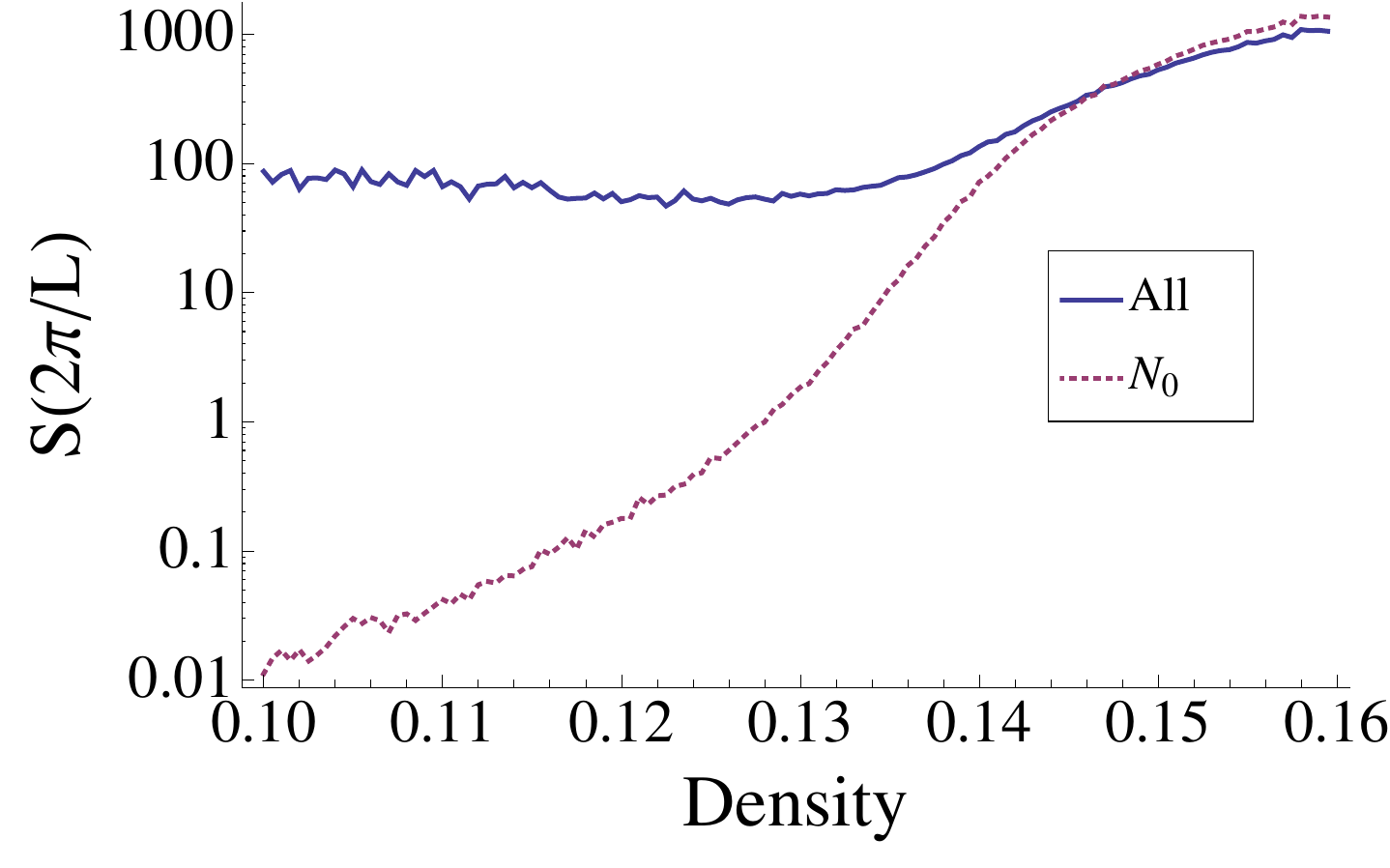}
\caption{(Color online) 
Behavior of $S(q\!=\!2\pi/L)$ and $S_{0}(q\!=\!2\pi/L)$ for $L=10000$.
 (a) $\Vmax=9$ and (b) $\Vmax=5$, both with $p=0.1$.\label{smallqplots}}
\end{figure}

To study the spatial correlations in $n_{0}(r)$, we examine
the static structure factor
\[
S_{0}(q) = \langle |\rho_{0}(q)|^2\rangle \qquad
\rho_{0}(q) = \sum_r n_{0}(r) e^{-iqr}\,.
\]
In Fig.~\ref{S0upturn} we show the magnitude of $S(q)$ and $S_{0}(q)$
at small values of $q$ for a several densities spanning the transition
to jamming for $\Vmax=9$ and $p=0.1$.  We see that most of the upturn in
$S(q)$ for small $q$ in the transition to jamming comes from correlations
in $n_{0}(r)$ of order less than a hundred lattice spacings or so.
The behavior for $\Vmax=5$ is similar, except that the width of the
peak in $S_0(q)$ (and $S(q)$) is much wider, indicating that the jammed
regions are significantly smaller at low $\Vmax$.
This behavior is also visible in Fig.~\ref{smallqplots}, where we
display the variation of $S(q)$ and $S_{0}(q)$ with density for the
longest wavelength $q=2\pi/L$ in our simulation.  
The behavior for $\Vmax=5$, shown in Fig.~\ref{smallqplots}(b) is very
similar to that seen for $\Vmax=9$, so we can conclude that the upturn
in $S(q)$ in the transition region is due to long-range correlations in
$n_0(r)$.  This is surprising, since we did not see finite-size effects
for $\Vmax=5$. We will see in the next section that it is the dynamics
and statistics of the jams that is different for the two situations.

\subsection{Jam Dynamics\label{Dynamics}}

\begin{figure}
\includegraphics[width=0.95\hsize]{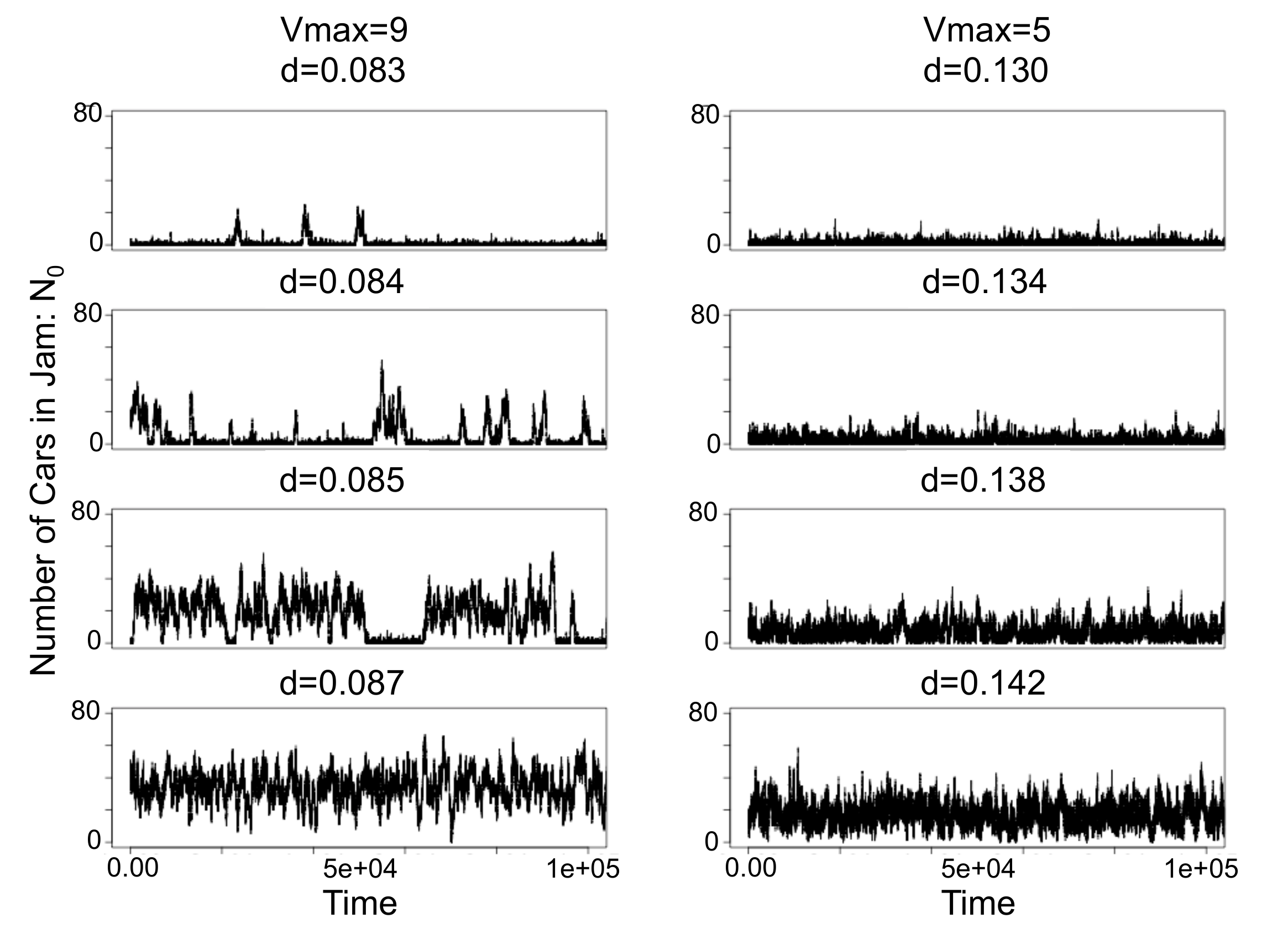}
\caption{
Time evolution of the number of cars in jam $N_{0}$ for a track
of length 5,000 for $\Vmax =9$ (left) and $\Vmax =5$ (right) and
$p=0.1$.\label{N0time}}
\end{figure}

To explore the difference between the low $\Vmax $ and high $\Vmax $
behavior, we looked at the time evolution of system at the transition
area.  Figure~\ref{N0time} shows how $N_{0}(t)$ behaves for different
densities for $\Vmax=9$ and $\Vmax=5$.  The densities shown are chosen
so that the system is initially in the pure free-flow phase, then early
in the transition region near the peak flux from Fig~\ref{flux}, then
late in the transition region, and finally in the jam phase.  In the
free flow phase, we see isolated fluctuations into the jam phase very
rarely for $\Vmax\!=\!9$, while these fluctuations are more frequent
in the $\Vmax\!=\!5$ simulations.  In the transition region close
to the peak in the flux shown in Fig.~\ref{flux}, the $\Vmax\!=\!9$
simulations still show isolated bursts of appearance of the jam phase,
while for $\Vmax\!=\!5$ the number of jammed vehicles is fluctuating
but always nonzero.

Since our order parameter $x_{0}$ for small $\Vmax$ might be showing
this behavior due to an inability to cleanly separate freely flowing
vehicles from jammed vehicles, we also used an order parameter that
examines second neighbor correlations.  Instead of just asking that the
distance to the car ahead be less than $\Vmax/2$, we consider three cars
in succession located at $r_n$, $r_{n+1}$, $r_{n+2}$ and require each
pair of cars to be spaced by less than $\Vmax/2$.
\[
\phi(r) =
\begin{cases}
1 & r_n = r \mbox{ and } \\
  & |r_{n+1}-r_n|<\Vmax/2 \mbox { and } \\
  & |r_{n+2}-r_{n+1}|<\Vmax/2 \\
0 & \mbox{otherwise\, ,}\\
\end{cases}
\]
The total number of cars with this condition is $\Phi_{0}=\sum_r \phi_(r)$.

\begin{figure}
\subfigure[]{\includegraphics[width=0.48\hsize]{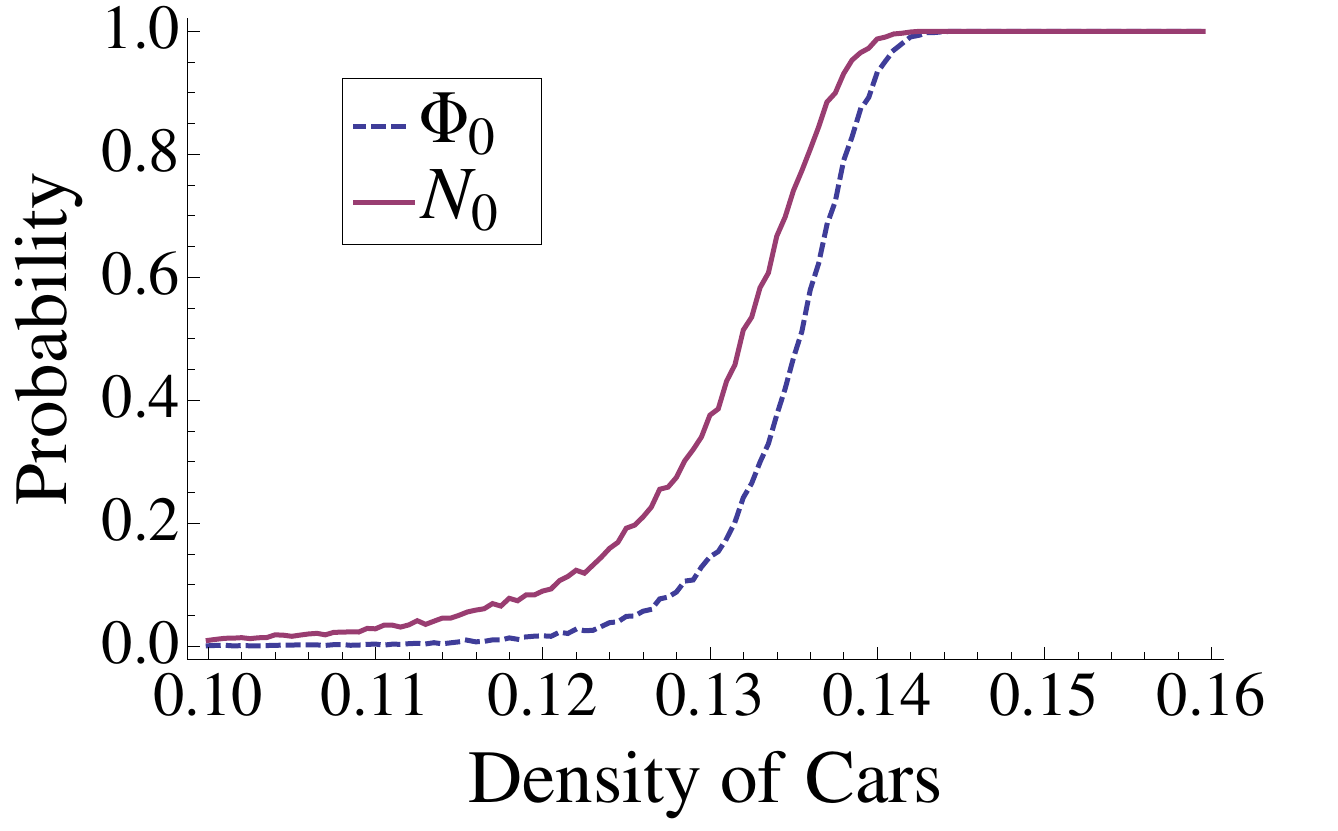} }
\subfigure[]{\includegraphics[width=0.48\hsize]{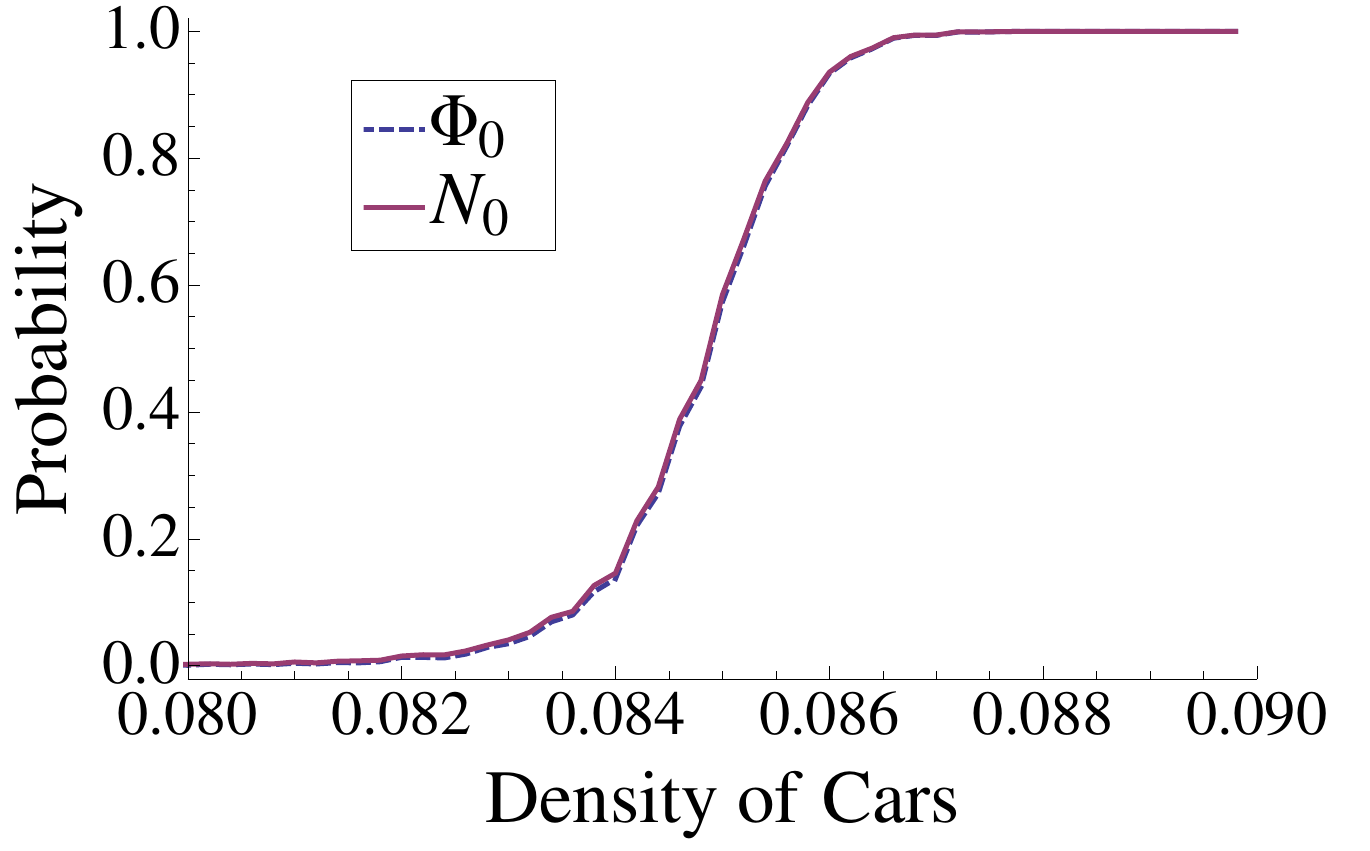} }
\caption{(Color online) 
Fraction of the time that the two order parameters $N_{0}$ and
$\Phi_{0}$ are nonzero for different densities for (a) $\Vmax=5$ and
(b) $\Vmax=9$.\label{N0Phi0}}
\end{figure}

We show in Fig.~\ref{N0Phi0} the fraction of the time the two order
parameters are nonzero for $\Vmax=5$ and $\Vmax=9$.  For $\Vmax=5$ both
order parameters are always nonzero after $d=0.115$, but that is actually
a density before the peak flux in Fig.~\ref{flux} occurs for $\Vmax=5$,
so the two order parameters become identical before the transition to
jams occurs.  For $\Vmax=9$, the two order parameters coincide and the
transition from nearly zero to unity is the density range where we see
finite-size effects.
So improving our definition of a jammed vehicle does not change the
conclusion that the nucleation of the jams for $\Vmax\alt 6$ is
qualitatively different than those for $\Vmax\agt 7$.

\begin{figure}
\includegraphics[width=0.95\hsize]{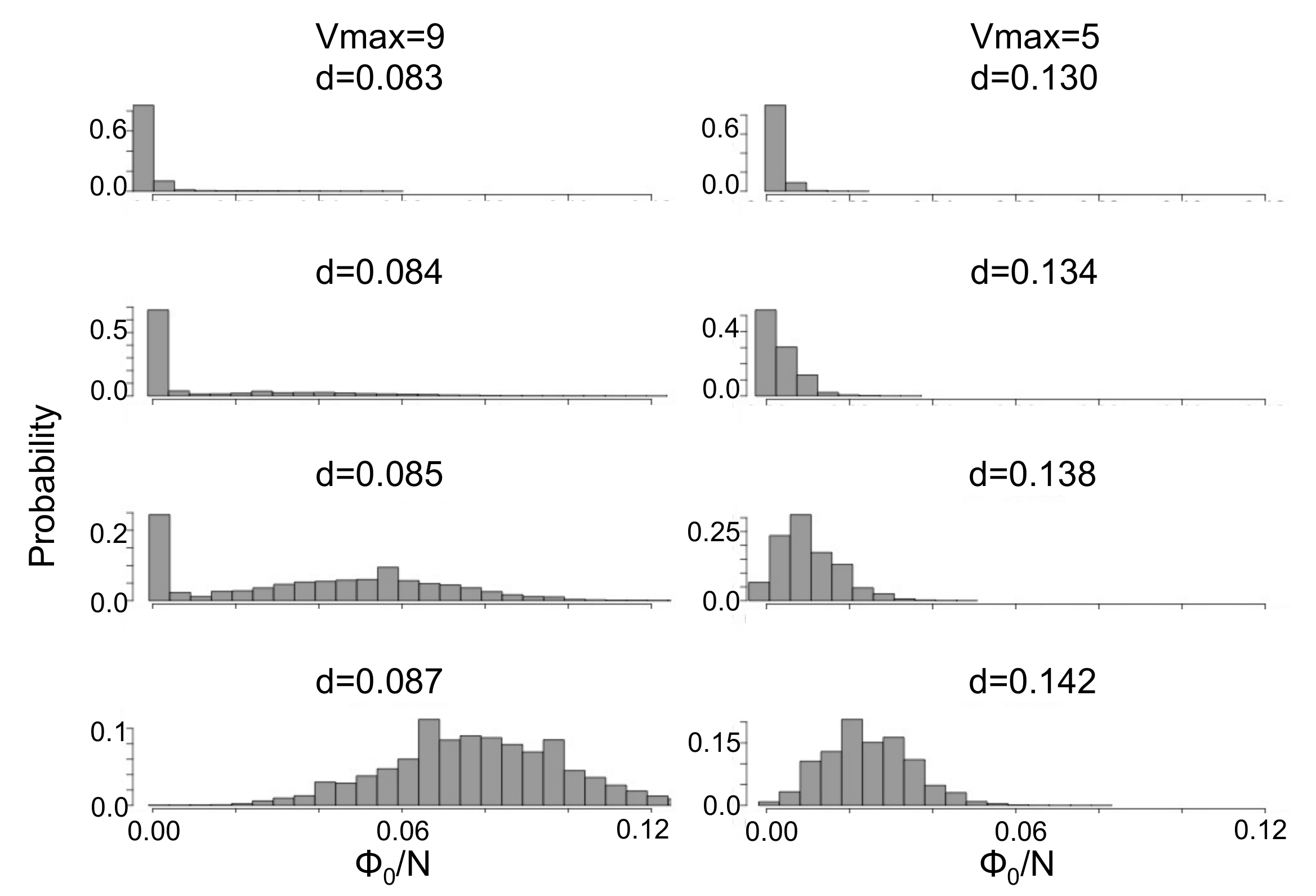}
\caption{
Histogram of $\Phi_{0}/N$ for the simulations show in
Fig.~\ref{N0time}.\label{N0hist}}
\end{figure}

The difference in behavior for different $\Vmax$ is also clearly visible
in histograms of the fraction of the cars in a jam, $\Phi_{0}/N$.
Figure~\ref{N0hist} shows the histograms for the same simulations shown
in Fig.~\ref{N0time}.  The peak at $\Phi_{0}/N=0$ is the vehicles in the
free-flow phase.  While the distributions for the free flow regime at
low density and the jammed regime are similar for both values of $\Vmax$,
they are clearly different in the transition regime.  For $\Vmax\!=\!9$
the jam phase appears as a distinct phase in the transition region,
while for $\Vmax\!=\!5$ this does not happen, with the distribution of
jammed cars growing smoothly out of the free-flow phase. 

\begin{figure}
\includegraphics[width=0.95\hsize]{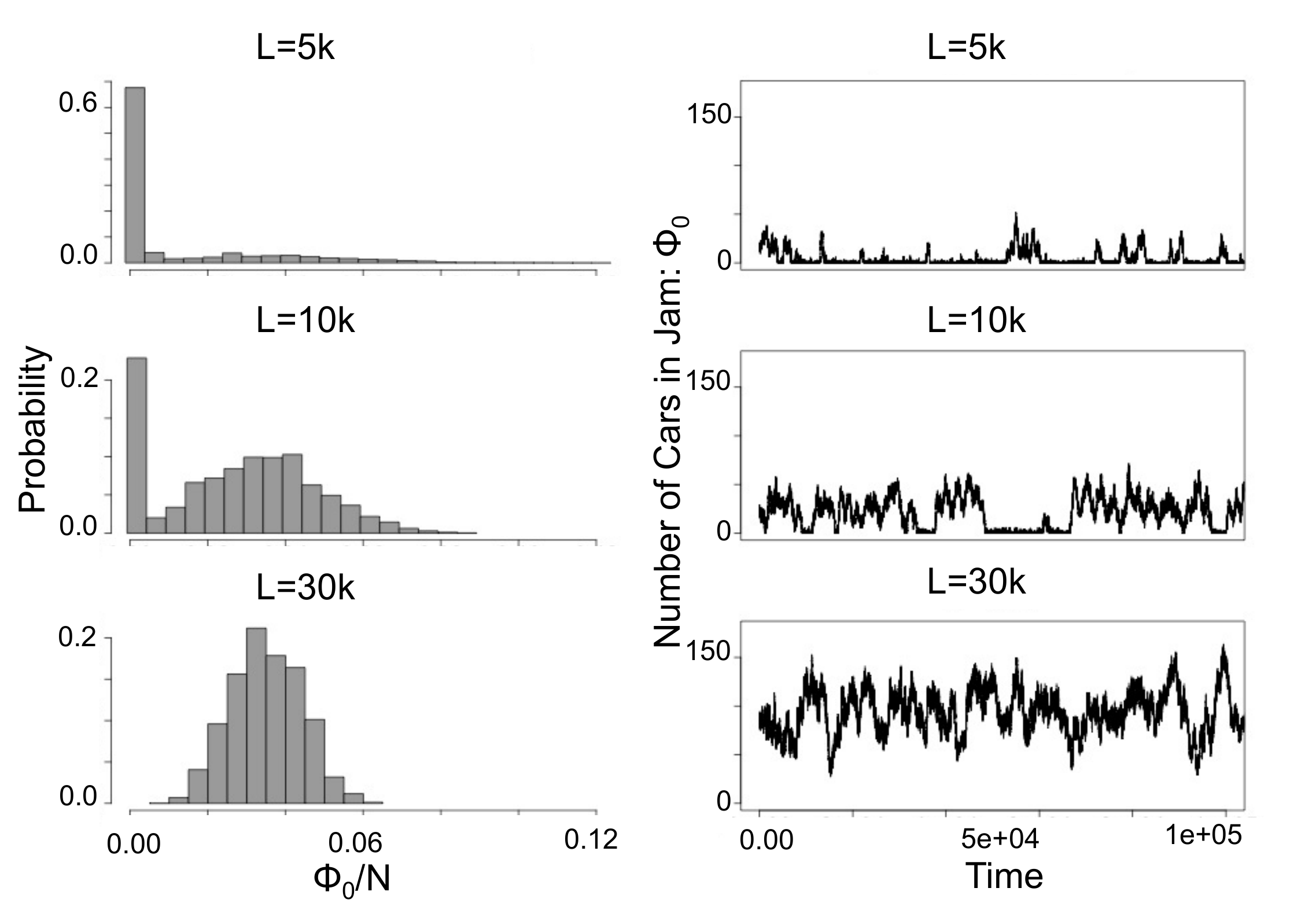}
\caption{Time evolution and histogram of $\Phi_{0}/N$
for systems with same density $d=0.084$ and different lengths.}
\label{N0histtime}
\end{figure}

Figure~\ref{N0histtime} shows how the length of the track affects
themselves in the time evolution $\Phi_{0}$ and in the distribution
of $\Phi_{0}$.  The density in all of the plots is the same, but the
histogram and the time evolution for the shortest track is characteristic
of the free phase regime.  For the intermediate track length, the
behavior is that of two-phase coexistence.  The longest track length
data show it to be in the jam phase.  Therefore we see that shorter
track lengths inhibit the transition to jamming and thus we expect to
see finite-size effects.

The transition region at high $\Vmax$ can thus be thought of as a
coexistence of cars condensed into jams and cars flowing freely.
The appearance of a single localized jam will reduce the density of
freely flowing cars elsewhere, and this effect is more significant for
shorter tracks.  Since the probability of creating a jammed region drops
as the density of freely flowing cars goes down, the appearance of one
jam inhibits the appearance of an additional one, stabilizing the dilute
gas of jams.

\begin{figure}
\subfigure[]{\includegraphics[width=0.48\hsize]{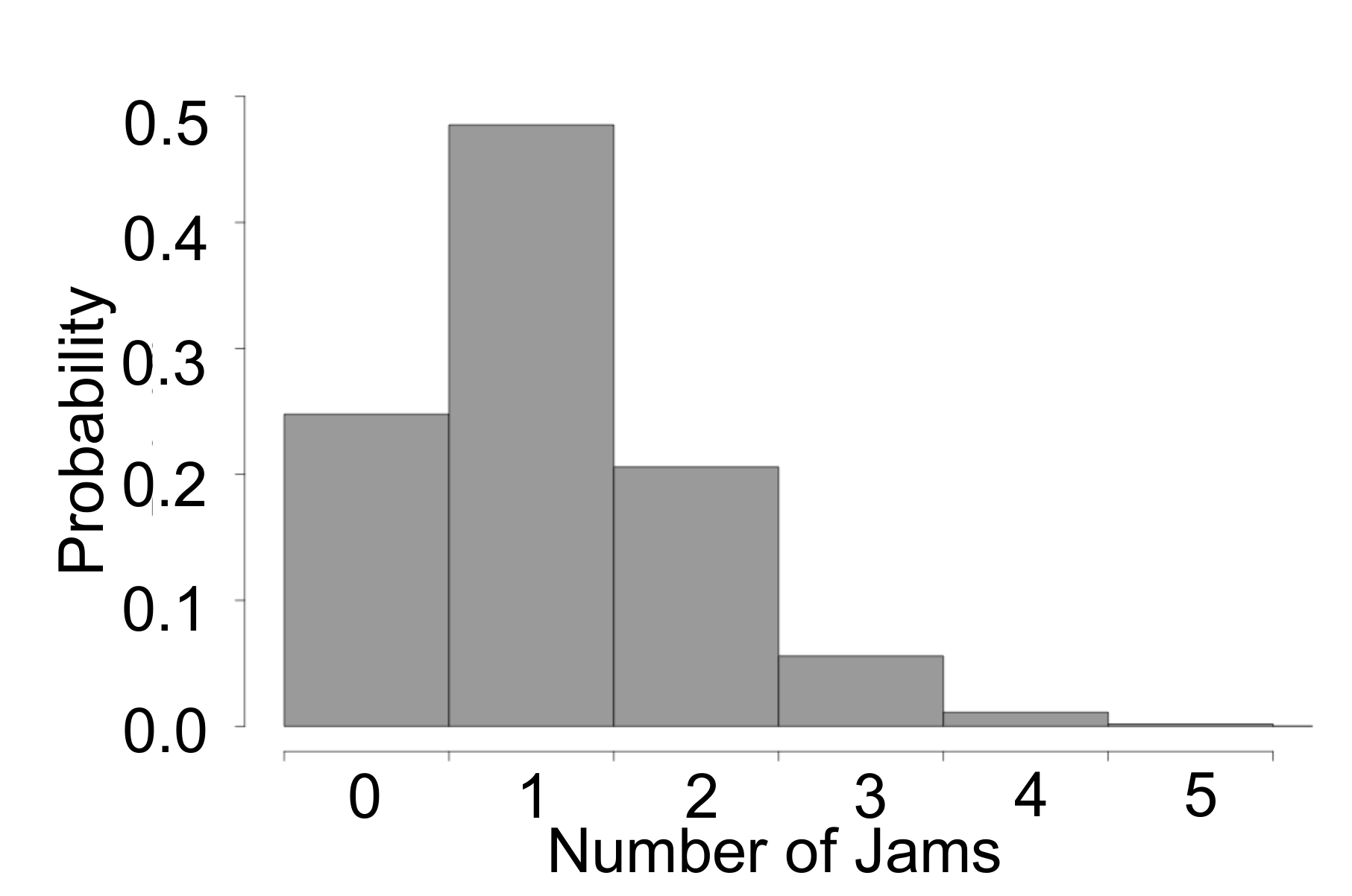} }
\subfigure[]{\includegraphics[width=0.48\hsize]{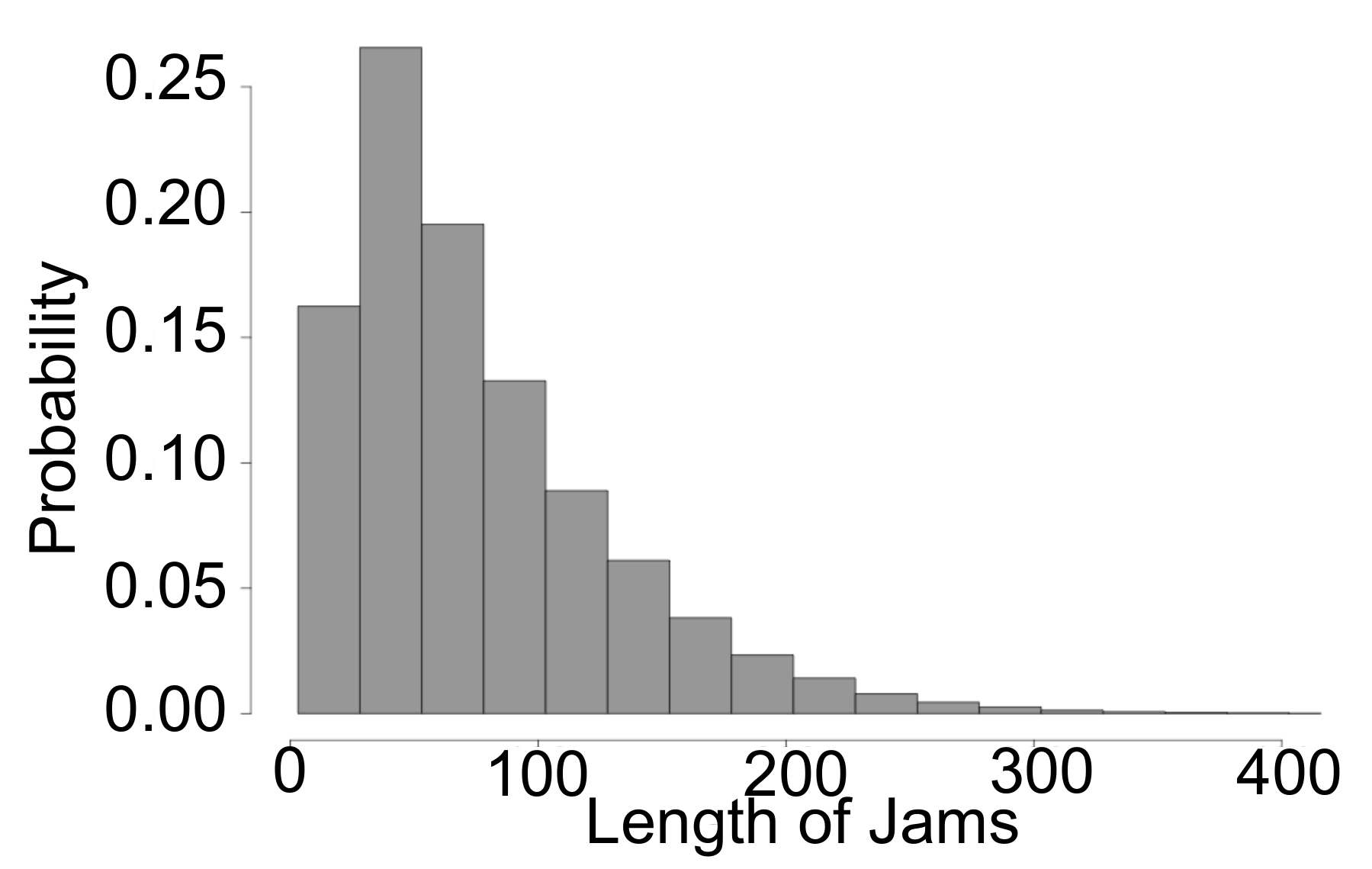} }
\caption{Statistics for 200,000 samples in a system of size $L=10,000$
for $\Vmax=9$, $p=0.1$ and $d=0.084$.  (a) Distribution of number
of jams (b) Distribution of jam lengths.\label{V9jamhisto}}

\end{figure}

This picture favors a fewer large jams rather many small jams.
Figure~\ref{V9jamhisto} shows a histogram of the number of jams for
$\Vmax=9$ at a density $d=0.084$, and also the distribution of jam
lengths.   The distribution is not Poisson, as we would expect for
independent events.  Instead, we see a marked tendency for one or two
large jams, with the probability of three or more jams greatly reduced.
We find that the jam would be of order 50 sites in length, in agreement
with the width of $S_{0}(q)$ for small $q$ seen in Fig.~\ref{S0upturn}.

\begin{figure}
\subfigure[]{\includegraphics[width=0.45\hsize]{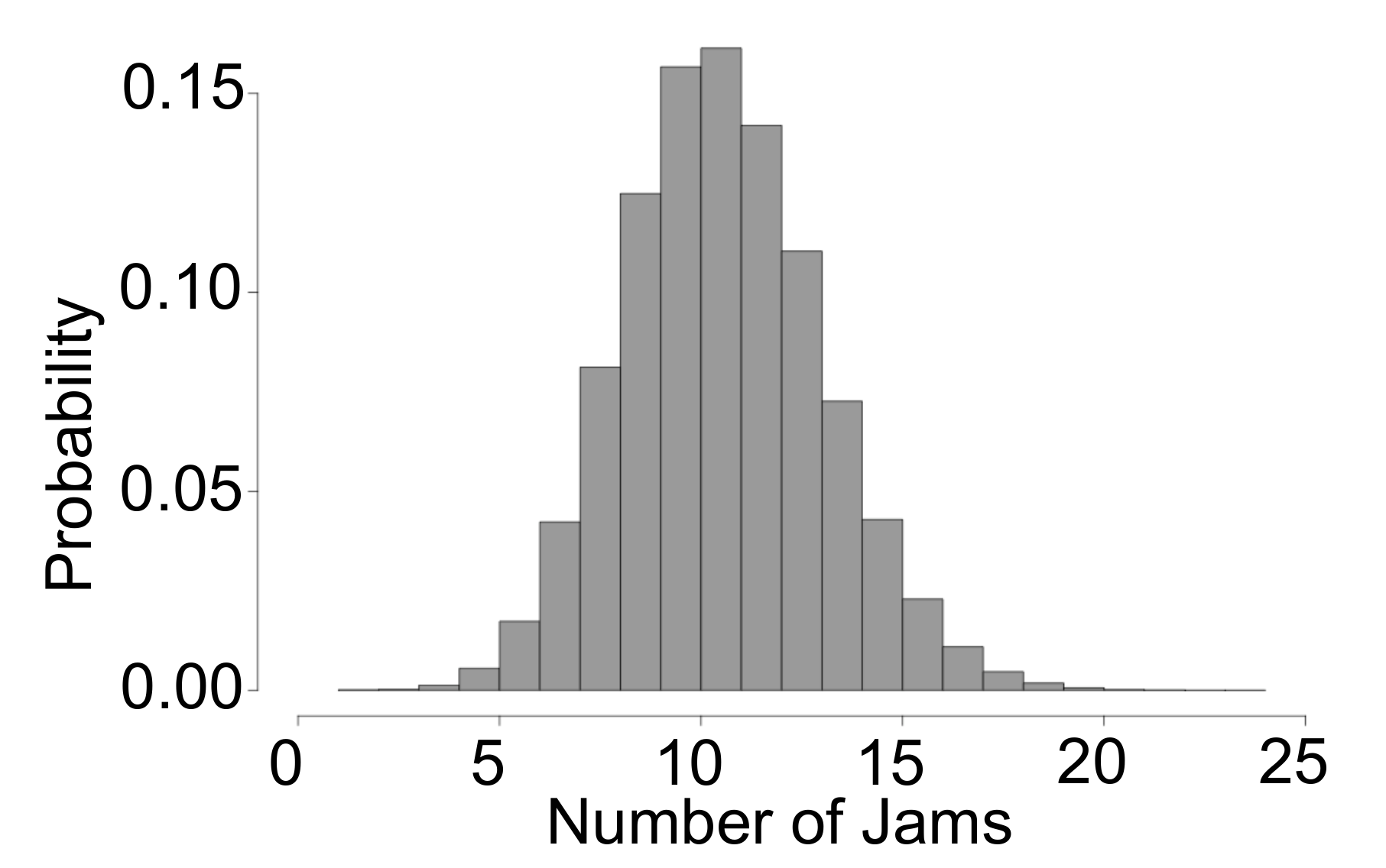} }
\subfigure[]{\includegraphics[width=0.45\hsize]{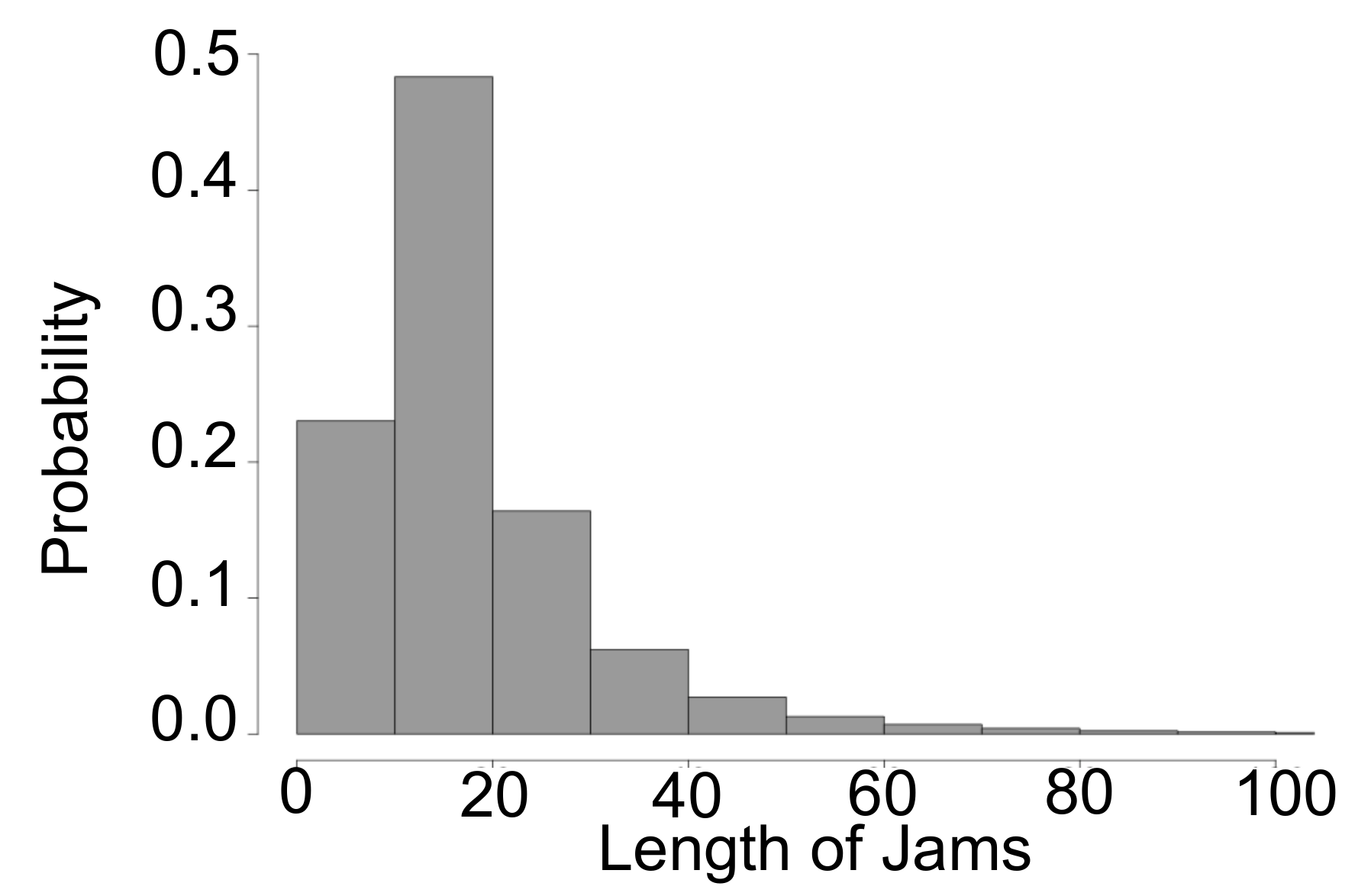} }
\caption{Statistics for 200,000 samples in a system of size $L=10,000$
for $\Vmax=5$, $p=0.1$ and $d=0.142$. 
(a) Distribution of number of jams and
(b) Distribution of jam lengths.\label{V5jamhisto}}
\end{figure}

For low $\Vmax$, the system does not break easily into tightly packed jams
and a lower density of freely flowing cars. The nucleation of one jam
does not depress the density of freely flowing cars sufficiently to inhibit
the formation of subsequent jams.  As a result, the low $\Vmax$ system
does not have a clear transition region where isolated jams appear,
and it is the isolated jams that are responsible for the finite-size
effects we see at higher $\Vmax$.  This is clearly seen in the distribution
of numbers of jams and jam lengths shown in Fig.~\ref{V5jamhisto}.  The
simulations were done at a density close to the peak flux for $\Vmax=5$.
The jams are smaller and more frequent than in the $\Vmax=9$ data of
Fig.~\ref{V9jamhisto}.

Once the nucleation of one jam does not significantly inhibit the
formation of a second one, we have reached the heavily jammed region and
the finite-size effects that appear at high $\Vmax$ disappear.  The mean
velocity and flux then follow the relations shown in Figs.~\ref{globalvel}
and \ref{flux} that are insensitive to the value of $\Vmax$.

\subsection{Coexistence at the Transition\label{analyticalt}}

\begin{figure}
\subfigure[]{\includegraphics[width=0.9\hsize]{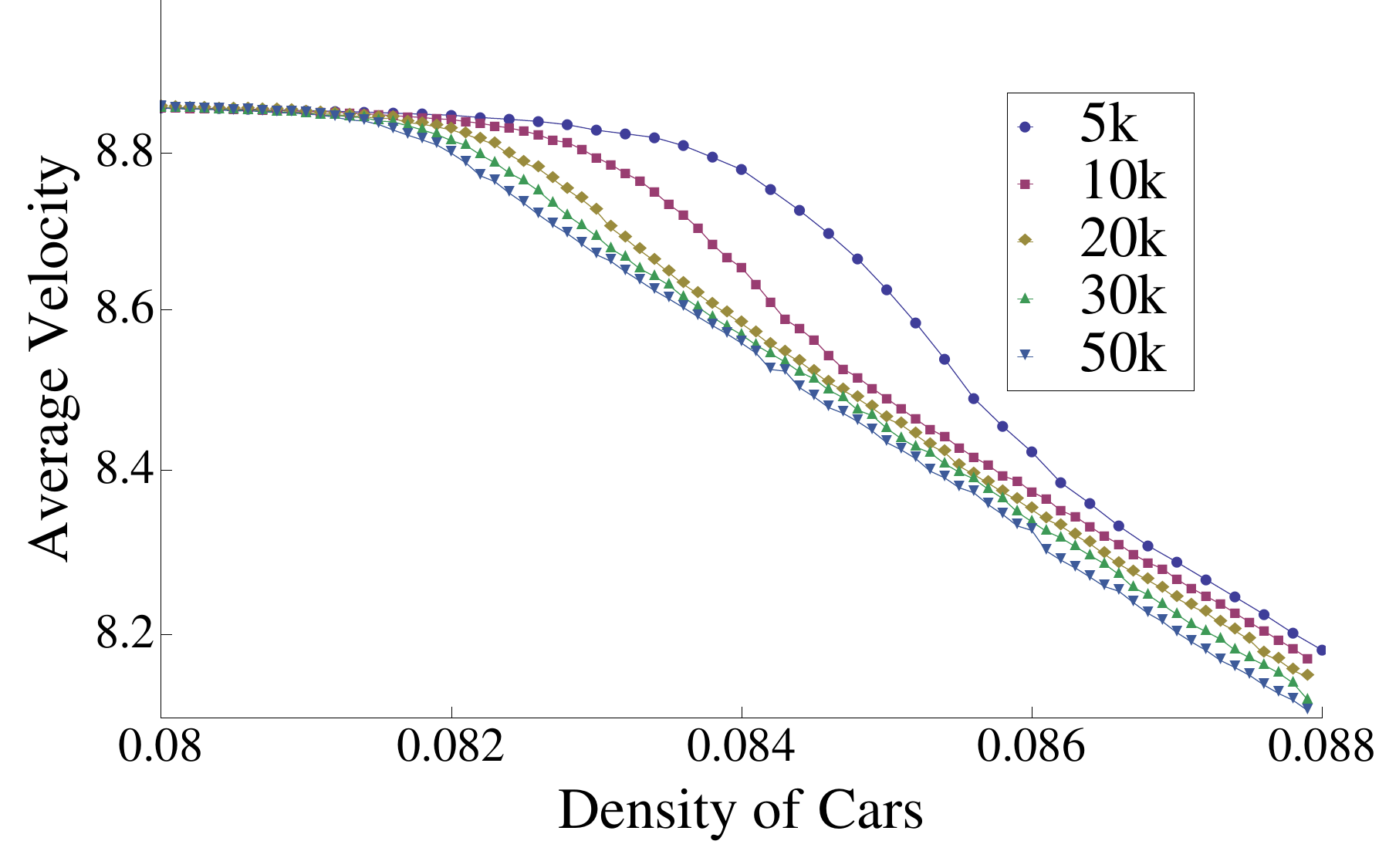} }
\subfigure[]{\includegraphics[width=0.9\hsize]{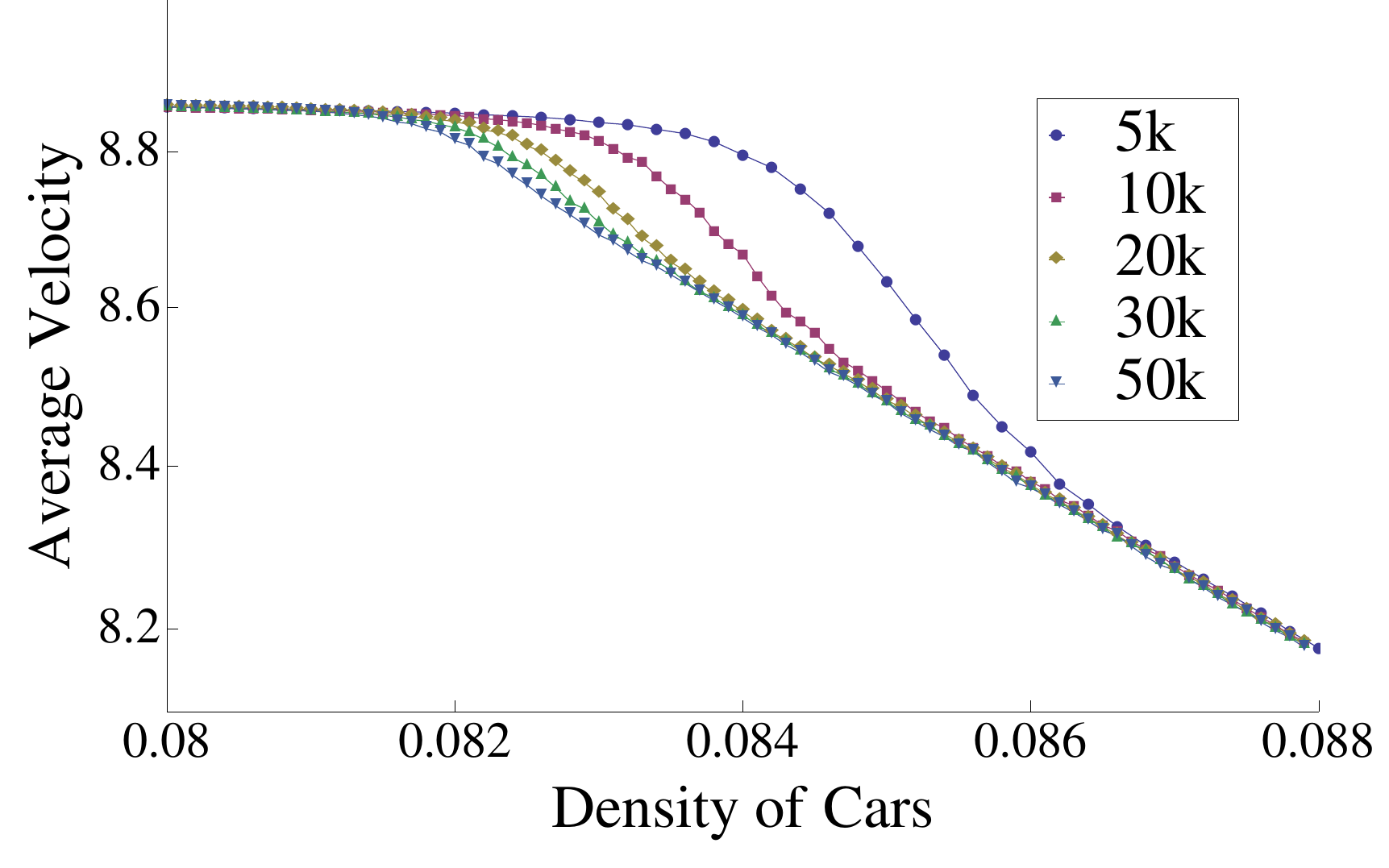} }

\caption{(Color online) (a) Calculated and (b) simulated average velocity
for different track lengths versus density of cars for $\Vmax =9$
and $p=0.1$.
\label{avev}}

\end{figure}

This picture of a few localized jams condensing out of the free flow
phase allows us to make a quantitative description of the finite-size
effects in the large $\Vmax$ simulations.  If we assume the jams consist
of cars traveling as fast as the gap rule would allow, then their mean
speed would be
\begin{equation}
V_{\text{J}} = \sum_{i<\Vmax/2} (i-1) \, P(i)\,,
\label{eqn:vj}
\end{equation}
and the fraction of the track occupied by the jams is
\begin{equation}
L_{\text{J}} = N\sum_{i<\Vmax/2} i \, P(i)\label{eqn:lj}\,.
\end{equation}
Since $N_{0}$ is the number of cars in a jam, then the density of free
cars is changed to
\begin{equation}
d_{\text{eff}} = \frac{N - N_{0}}{L - L_{\text{J}}}\label{eqn:deff}\,,
\end{equation}
We then expect that the mean velocity of the mixture would be
\begin{equation}
V_{\text{tot}} = \frac{N_{0}}{N} V_{\text{J}}
+ \left(1-\frac{N_{0}}{N}\right)V_{\text{F}}(d_{\text{eff}})\,,
\label{eqn:vbar}
\end{equation}
where $V_{\text{F}}(d_{\text{eff}})$ is the flow velocity
in the free flow phase taken from the average velocity plot
(Fig.~\ref{globalvel}) at an effective density $d_{\text{eff}}$ given by
Eq.~(\ref{eqn:deff}).  We evaluate Eqs.~(\ref{eqn:vj}), (\ref{eqn:lj}),
and $V_{\text{F}}(d_{\text{eff}})$ directly from our simulations to
find $V_{\text{tot}}$.  Figure~\ref{avev} shows the result of this for
several track lengths.  The agreement with the simulations is excellent
in the transition region, and underestimates the mean velocity at higher
density where this simple picture of two phase coexistence breaks down.

\section{Conclusions\label{Conc}}

We have shown that the interactions in the NS model for a dilute
gas of vehicles can be described as weakly interacting except for a
repulsive interaction of range $\approx \Vmax$.  The gap distribution
can be found from solving the kinetic equations keeping only these
pair interactions.  Spatial correlations among cars further apart can
be described quantitatively through an Ornstein-Zernicke relation.

As the density is raised to the point where jams form, we see that the
nearest neighbor interactions fail to describe the density correlations.
The jam formation results in a peak in the gap distribution $P(r)$ for
small $r$ arising from jammed vehicles.  At the same time, the structure
factor $S(q)$ shows a significant upturn for small $q$, indicating the
onset of long range correlations.  These new correlations are not the
result of including just second or third neighbor correlations.

Several authors~\cite{Lubeck1998,Chowdhury1998,Roters1999} have noticed
that the transition to jams can be described as the nucleation of
isolated jams in a background of freely flowing cars.  Our work shows
that the characteristics of the transition depend on the value of $\Vmax$.
In this regard, $\Vmax$ plays the role of the number of degrees of freedom
(possible velocities) for each object, much like the role of the number
of components of a spin variable in an equilibrium system.

Systems with $\Vmax \agt 6$ show a transition with an intermediate
phase that exhibits significant finite size effects.  We attribute these
effects to the existence of large isolated jams that coexist with the
free flow phase in this intermediate regime.  These large jams act to
segregate vehicles and keep the free flow phase stable.  The finite-size
scaling analysis shows that these long range correlations appear to be
universal, with scaling exponents that depend on $\Vmax$.  We are able
to quantitatively fit the overshoot seen in the vehicle flux in a finite
size system by accounting for the segregation of the vehicles into jams.
While our choice for an order parameter was convenient in terms of
getting good statistics, we have shown that other order parameters that
have been used also show these finite size effects.

For $\Vmax\alt 6$ we are unable to separate the vehicles into two phases
and finite size effects in the transition region are either absent or
extremely small.  However, both high and low values of $\Vmax$ show an
upturn in the structure factor, indicating that the regions of large
correlated motion, even if they cannot be cleanly denoted as jams,
are responsible for the long range correlations in $S(q)$.

Since we see a smooth growth of the order parameter with density,
a peak in the susceptibility $\chi_4$ and long range correlations in
the density, we would characterize the transition to jamming as a second
order transition.  However, unlike an equilibrium second order transition,
the finite size effects arise not from long range correlations but rather
because the jam formation lowers the density and delays the onset to
multiple jams.

\appendix*

\section{Kinetic Model for Dilute Traffic\label{kinetic}}

In the dilute limit, the interaction between cars produced by the gap
rule only applies to a pair of vehicles at a time, and simultaneous
interactions among a triple of adjacent vehicles are rare.  We define
a distribution function $f(v,g)$ as the probability that a car has a
velocity $v$ and the gap to the vehicle ahead is $g$.

From this
distribution we can calculate the velocity distribution $P_v(v)$ as
\begin{equation}
P_v(v)=\sum\limits_{g=0}^{L-1} f(v,g)\,.
\label{veld}
\end{equation}
We can also calculate the distribution of gaps $\Delta(g)$ via
\begin{equation}
\Delta(g)=\sum_{v=0}^{\Vmax} f(v,g)\,,
\label{gapd}
\end{equation}
from which we can find the nearest neighbor distribution $P(r)$ from
the relation $P(r)=\Delta(r-1)$.  

Each of the 4 rules of the NS model alters the form of $f(v,g)$.  We find
it simplest to examine $f(v,g)$ right after the velocity updates and
before the position update.  This is tantamount to assigning the position
update as the first step instead of the last.

The position update
rule produces an altered distribution $\hat{f}(v,g)$ via
\begin{equation}
\hat{f}(v,g) = \sum_{u=0}^m P_v(u) f(v,g+v-u)_t\,,
\label{gap:evol}
\end{equation}
where for convenience we have denoted $\Vmax$ as $m$, since it will appear
frequently in this section.
Since we are ignoring triple correlations, the speed distribution of the
vehicle ahead is $P_v(v)$ from Eq.~(\ref{veld}).  
The velocity update rules then alter the $\hat{f}(v,g)$ distribution.  For
gaps greater than or equal to $\Vmax$
the rules yield
\begin{eqnarray}
f(m,g) &=& (1-p)\left[\hat{f}(m,g)+\hat{f}(m\!-\!1,g)\right] \nonumber \\
f(m\!-\!1,g)
       & = & p\left[\hat{f}(m,g)+\hat{f}(m\!-\!1,g)\right] + \nonumber \\
       &   & + (1-p) \hat{f}(m\!-\!2,g) \nonumber \\
f(v,g) &=& p\hat{f}(v,g) + (1-p) \hat{f}(v-1,g) \nonumber \\
       & & \qquad \qquad v=1\dots m-2 \nonumber \\
f(0,g) & = & p\hat{f}(0,g)  \,,
\end{eqnarray}
while for gaps smaller than $\Vmax$ we have
\begin{eqnarray}
f(v,g)_{t+1} &=&
0 \qquad  v=g+1\dots m \nonumber \\
f(g,g)_{t+1} &=&
(1-p)\left[\sum_{u=g-1}^m \hat{f}(u,g)\right] \nonumber \\
f(g\!-\!1,g)_{t+1} &=& p\left[\sum_{u=g-1}^m \hat{f}(u,g)\right]
                  + (1-p) \hat{f}(g\!-\!2,g) \nonumber \\
f(v,g)_{t+1} &=& p\hat{f}(v,g) + (1-p) \hat{f}(v-1,g) \nonumber \\
             &   & \qquad v=1\dots g-2 \nonumber \\
f(0,g)_{t+1} &=& p\hat{f}(0,g)\,.
\end{eqnarray}

Our simulations show that the 3 velocity update steps for dilute traffic
rapidly create a local equilibrium in the velocity distribution where the
vehicle is moving at the highest speed it can with probability $1-p$,
or at the next to highest speed with probability $p$.  For $g\ge m$,
the highest speed is $\Vmax$ and so the distribution is
\begin{eqnarray}
f(v,g) =
\begin{cases}
(1-p) \Delta(g) & v=m \\
p \Delta(g) & v=m-1\\
0 & v=0\dots m-2 \,,\\
\end{cases}
\label{fdist1}
\end{eqnarray}
while for gaps less than $\Vmax$ we have
\begin{eqnarray}
f(v,g) =
\begin{cases}
0 & v=g+1\dots m \\
(1-p) \Delta(g) & v=g \\
p \Delta(g) & v=g-1\\
0 & v=0\dots g-2\,.\\
\end{cases} 
\label{fdist2}
\end{eqnarray}
Deviations from this distribution relax exponentially
as $p^n$ after $n$ steps. 

The leading order correction that arises from the gap rule occurs when
a faster vehicle catches up to a slower vehicle so that the gap between
them is $\Vmax-1$.  The gap rule then limits the speed of the car behind
to $\Vmax-1$, causing it to spend a fraction of its time at a speed
of $\Vmax-2$.  The fraction of vehicles that do that is $P_v(\Vmax-2)
= p\Delta(\Vmax-1)$.

\begin{figure}%[!ht]
\subfigure[]{\includegraphics[width=0.48\hsize]{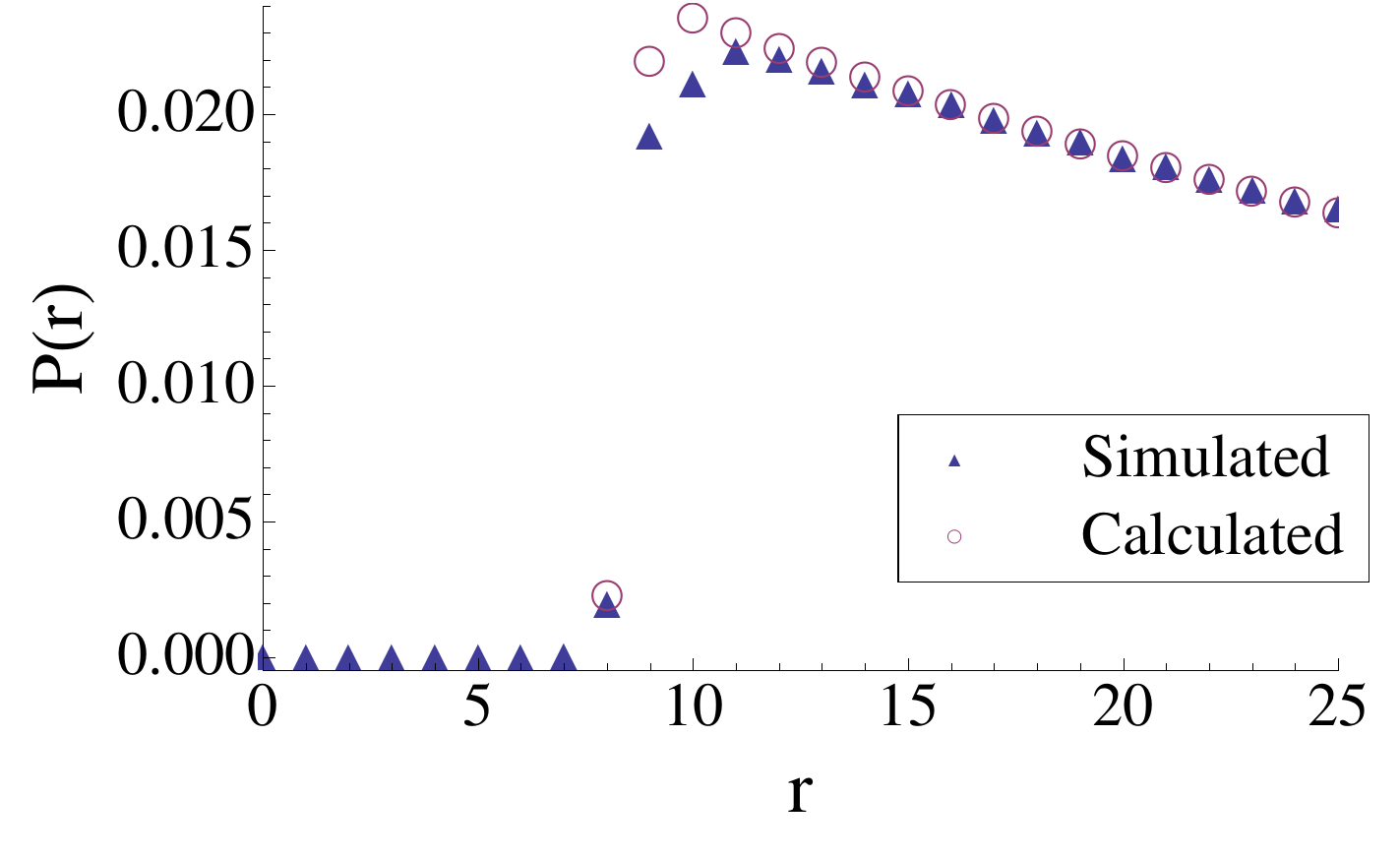} }
\subfigure[]{\includegraphics[width=0.48\hsize]{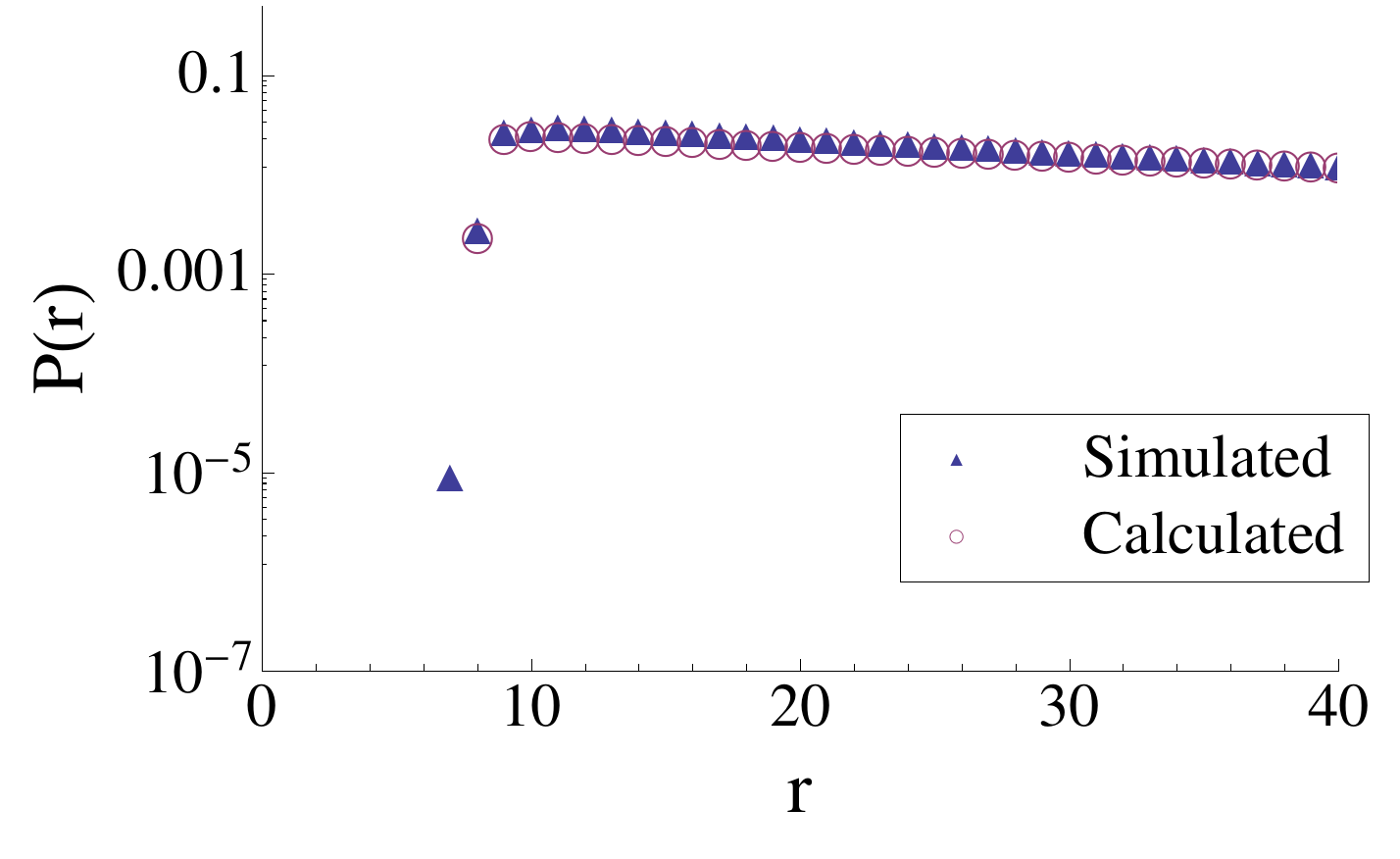} }
\caption{(Color online) 
Plots of the simulations and Eq.~(\ref{steadystate}) for
$V=9$ and $p=0.1$ at a density of 0.02 on (a) a regular scale and (b)
a semilog scale.\label{finacc}}
\end{figure} 

\begin{figure}%[!ht]
\subfigure[]{\includegraphics[width=0.48\hsize]{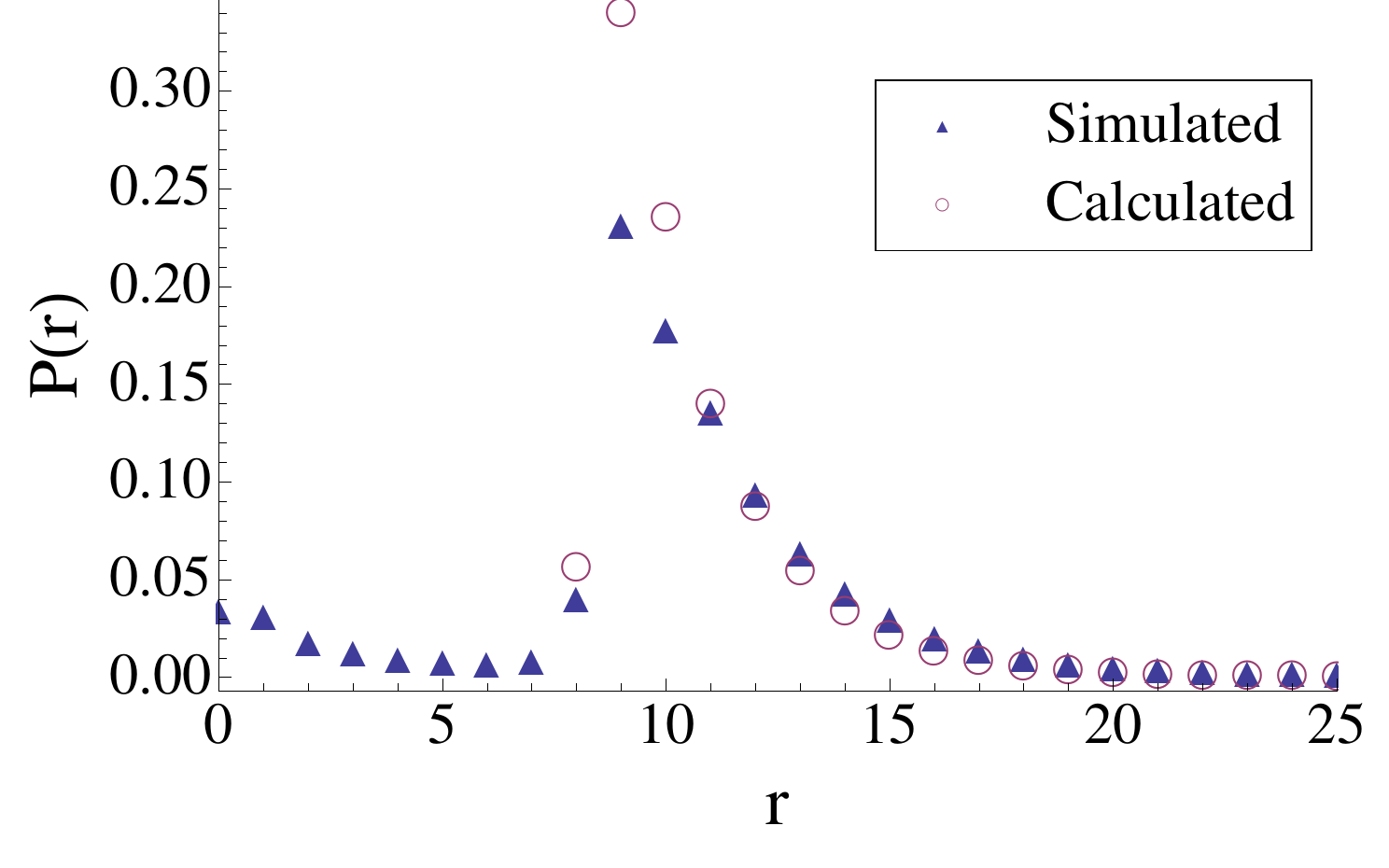} }
\subfigure[]{\includegraphics[width=0.48\hsize]{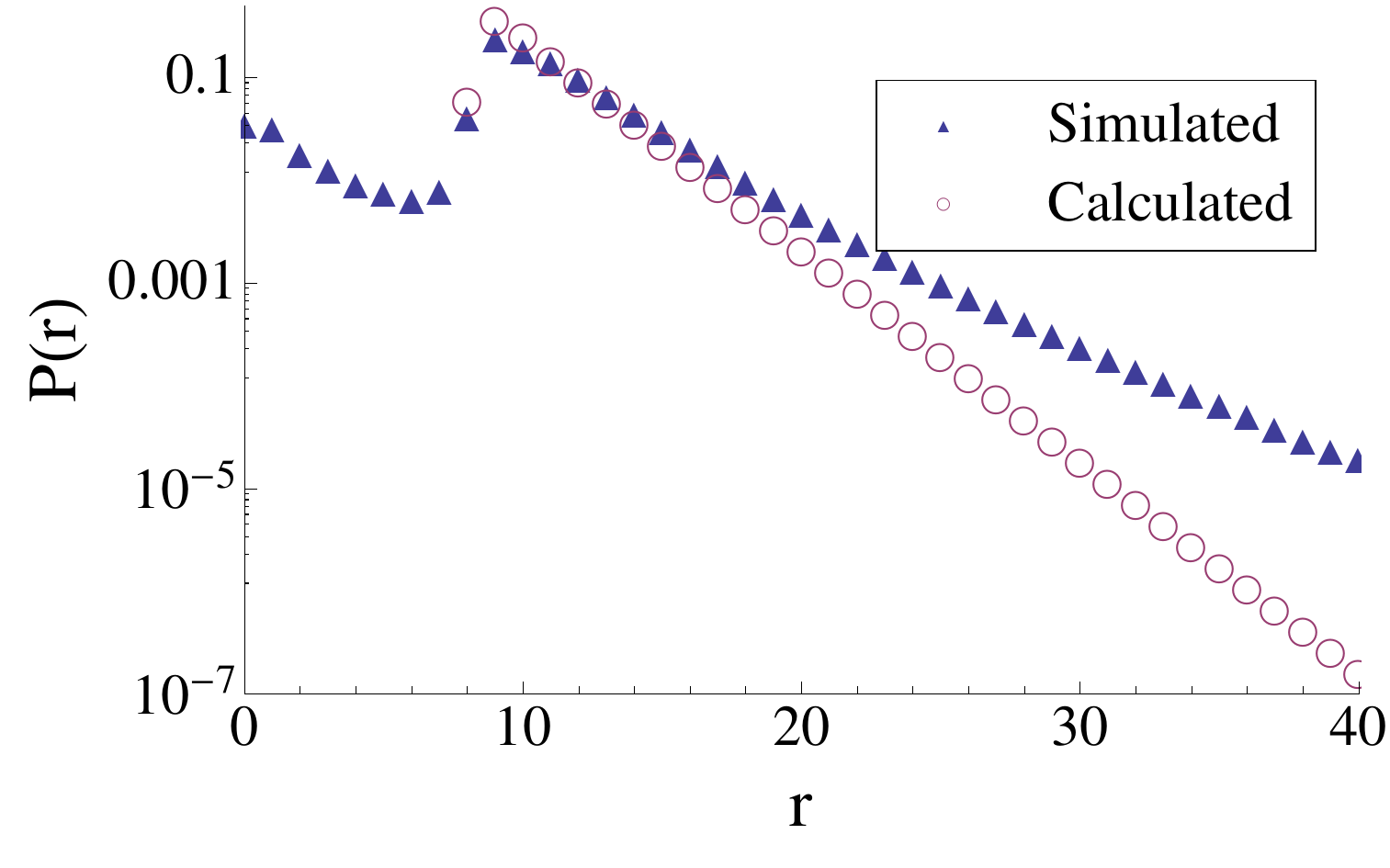} }
\caption{(Color online)
Comparison of the simulations and the analytic prediction
at a higher density of 0.09 for $\Vmax=9$ and $p=0.1$.
\label{finacchigh}}
\end{figure} 

The speed distribution from Eq.~(\ref{veld}) is then
\begin{eqnarray}
P_v(m\!-\!2)&=&\alpha p   \nonumber\\
P_v(m\!-\!1)&=& (1-\alpha) p +\alpha (1-p) \nonumber\\
P_v(m) &=&  (1-\alpha) (1-p)\,,
\label{vdis}
\end{eqnarray}
where $\alpha = \Delta(m\!-\!1)$ is the fraction of cars with a gap
of $\Vmax-1$.
Putting this speed distribution into Eq.~(\ref{gap:evol}) with the
assumed distribution for $f(v,g)$ given by Eqs.~(\ref{fdist1})
and (\ref{fdist2}), we can produce
an evolution equation for the
gap distribution of the form
\[
\Delta(g)_{t+1} = \sum_{g'} \Gamma(g'\to g) \Delta(g')_{t}\,.
%\label{delta}
\]
We find for $g\ge m+2$ that
\begin{eqnarray}
\Delta(g)_{t+1} - \Delta(g)_t
& = & A ( \Delta(g\!-\!1)_t-\Delta(g)_t) + \nonumber \\
&   & B (\Delta(g\!+\!1)_t-\Delta(g)_t) + \nonumber \\
&   & C (\Delta(g\!+\!2)_t-\Delta(g)_t) \nonumber \\
& \equiv & \Phi[\Delta(g)]\,,
\label{general}
\end{eqnarray}
where $A=p'(1-\alpha)$, $B=p'+\alpha(1-3p')$ and $C=\alpha p'$,
with $p'=p(1-p)$. 
The evolution equations for smaller gaps are then
\begin{eqnarray}
\Delta(m\!+\!1)_{t+1}\!-\!\Delta(m\!+\!1)_t & = &
\Phi[\Delta(m\!+\!1)] + A\Delta(m\!-\!1)_t \nonumber \\
\Delta(m)_{t+1}\!-\!\Delta(m)_t & = &
\Phi[\Delta(m)] \nonumber \\
 & + & (1\!-\!3p'\!+\!\alpha(1\!-\!2p'))\Delta(m\!-\!1)_t \nonumber \\
\Delta(m\!-\!1)_{t+1} & = & \alpha \,.
\label{leftend}
\end{eqnarray}

In the continuum limit and for $g>m+1$, Eq.~(\ref{general}) becomes a
drift-diffusion Fokker-Plank equation of the form
\begin{equation}
\frac{\partial\Delta}{\partial t} = 
\alpha \frac{\partial\Delta}{\partial g}
+ (p(1-p)+\alpha/2) \frac{\partial^2\Delta}{\partial g^2}\,,
\label{dif1}
\end{equation}
for which the steady-state solution is of the form
\begin{equation}
\Delta(g) \propto \exp\left( - \frac{\alpha}{p(1-p)+\alpha/2}\, g\right)\,.
\label{steadystate}
\end{equation}

If we solve Eqs.~(\ref{general}) and (\ref{leftend}) and compare them
to our simulations, we see from Fig.~\ref{finacc} that the agreement
is excellent except near the peak of the distribution.  Since both
distributions are normalized, the error at the peak results in slightly
different slopes for large $g$.  At higher densities, the agreement is
not as good.  Figure~\ref{finacchigh} shows the analytic description
predicts the position of the peak at $g=\Vmax$, but the presence of the
second jam phase in the simulations alters the distribution.

\bibliography{article}

\begin{thebibliography}{31}%
\makeatletter
\providecommand \@ifxundefined [1]{%
 \@ifx{#1\undefined}
}%
\providecommand \@ifnum [1]{%
 \ifnum #1\expandafter \@firstoftwo
 \else \expandafter \@secondoftwo
 \fi
}%
\providecommand \@ifx [1]{%
 \ifx #1\expandafter \@firstoftwo
 \else \expandafter \@secondoftwo
 \fi
}%
\providecommand \natexlab [1]{#1}%
\providecommand \enquote  [1]{``#1''}%
\providecommand \bibnamefont  [1]{#1}%
\providecommand \bibfnamefont [1]{#1}%
\providecommand \citenamefont [1]{#1}%
\providecommand \href@noop [0]{\@secondoftwo}%
\providecommand \href [0]{\begingroup \@sanitize@url \@href}%
\providecommand \@href[1]{\@@startlink{#1}\@@href}%
\providecommand \@@href[1]{\endgroup#1\@@endlink}%
\providecommand \@sanitize@url [0]{\catcode `\\12\catcode `\$12\catcode
  `\&12\catcode `\#12\catcode `\^12\catcode `\_12\catcode `\%12\relax}%
\providecommand \@@startlink[1]{}%
\providecommand \@@endlink[0]{}%
\providecommand \url  [0]{\begingroup\@sanitize@url \@url }%
\providecommand \@url [1]{\endgroup\@href {#1}{\urlprefix }}%
\providecommand \urlprefix  [0]{URL }%
\providecommand \Eprint [0]{\href }%
\providecommand \doibase [0]{http://dx.doi.org/}%
\providecommand \selectlanguage [0]{\@gobble}%
\providecommand \bibinfo  [0]{\@secondoftwo}%
\providecommand \bibfield  [0]{\@secondoftwo}%
\providecommand \translation [1]{[#1]}%
\providecommand \BibitemOpen [0]{}%
\providecommand \bibitemStop [0]{}%
\providecommand \bibitemNoStop [0]{.\EOS\space}%
\providecommand \EOS [0]{\spacefactor3000\relax}%
\providecommand \BibitemShut  [1]{\csname bibitem#1\endcsname}%
\let\auto@bib@innerbib\@empty
%</preamble>
\bibitem [{\citenamefont {Lighthill}\ and\ \citenamefont
  {Whitham}(1955)}]{FluidDyn}%
  \BibitemOpen
  \bibfield  {author} {\bibinfo {author} {\bibfnamefont {M.~J.}\ \bibnamefont
  {Lighthill}}\ and\ \bibinfo {author} {\bibfnamefont {G.~B.}\ \bibnamefont
  {Whitham}},\ }\href@noop {} {\bibfield  {journal} {\bibinfo  {journal}
  {Proc.\ Roy.\ Soc.}\ }\bibinfo {series} {A},\ \textbf {\bibinfo {volume}
  {229}},\ \bibinfo {pages} {317} (\bibinfo {year} {1955})}\BibitemShut
  {NoStop}%
\bibitem [{\citenamefont {Prigogine}\ and\ \citenamefont
  {Herman}(1971)}]{Boltz}%
  \BibitemOpen
  \bibfield  {author} {\bibinfo {author} {\bibfnamefont {I.}~\bibnamefont
  {Prigogine}}\ and\ \bibinfo {author} {\bibfnamefont {R.}~\bibnamefont
  {Herman}},\ }\href@noop {} {\emph {\bibinfo {title} {Kinetic Theory of
  Vehicular Traffic}}}\ (\bibinfo  {publisher} {Elsevier},\ \bibinfo {year}
  {1971})\BibitemShut {NoStop}%
\bibitem [{\citenamefont {Chowdhury}(2000)}]{Chowdhury2000}%
  \BibitemOpen
  \bibfield  {author} {\bibinfo {author} {\bibfnamefont {D.}~\bibnamefont
  {Chowdhury}},\ }\href {\doibase 10.1016/S0370-1573(99)00117-9} {\bibfield
  {journal} {\bibinfo  {journal} {Phys.\ Rep.}\ }\textbf {\bibinfo {volume}
  {329}},\ \bibinfo {pages} {199} (\bibinfo {year} {2000})}\BibitemShut
  {NoStop}%
\bibitem [{\citenamefont {Helbing}(2001)}]{Helbing2001}%
  \BibitemOpen
  \bibfield  {author} {\bibinfo {author} {\bibfnamefont {D.}~\bibnamefont
  {Helbing}},\ }\href {\doibase 10.1103/RevModPhys.73.1067} {\bibfield
  {journal} {\bibinfo  {journal} {Rev.\ Mod.\ Phys.}\ }\textbf {\bibinfo
  {volume} {73}},\ \bibinfo {pages} {1067} (\bibinfo {year}
  {2001})}\BibitemShut {NoStop}%
\bibitem [{\citenamefont {Nagatani}(2002)}]{Nagatani2002}%
  \BibitemOpen
  \bibfield  {author} {\bibinfo {author} {\bibfnamefont {T.}~\bibnamefont
  {Nagatani}},\ }\href {\doibase 10.1088/0034-4885/65/9/203} {\bibfield
  {journal} {\bibinfo  {journal} {Rep.\ Prog.\ Phys.}\ }\textbf {\bibinfo
  {volume} {65}},\ \bibinfo {pages} {1331} (\bibinfo {year}
  {2002})}\BibitemShut {NoStop}%
\bibitem [{\citenamefont {Nagel}\ and\ \citenamefont
  {Schreckenberg}(1992)}]{Nagel1992}%
  \BibitemOpen
  \bibfield  {author} {\bibinfo {author} {\bibfnamefont {K.}~\bibnamefont
  {Nagel}}\ and\ \bibinfo {author} {\bibfnamefont {M.}~\bibnamefont
  {Schreckenberg}},\ }\href {\doibase 10.1051/jp1:1992277} {\bibfield
  {journal} {\bibinfo  {journal} {J.\ Phys. I}\ }\textbf {\bibinfo {volume}
  {2}},\ \bibinfo {pages} {2221} (\bibinfo {year} {1992})}\BibitemShut
  {NoStop}%
\bibitem [{\citenamefont {Nagel}\ and\ \citenamefont
  {Paczuski}(1995)}]{Nagel1995}%
  \BibitemOpen
  \bibfield  {author} {\bibinfo {author} {\bibfnamefont {K.}~\bibnamefont
  {Nagel}}\ and\ \bibinfo {author} {\bibfnamefont {M.}~\bibnamefont
  {Paczuski}},\ }\href {\doibase 10.1103/PhysRevE.51.2909} {\bibfield
  {journal} {\bibinfo  {journal} {Phys.\ Rev.\ E}\ }\textbf {\bibinfo {volume}
  {51}},\ \bibinfo {pages} {2909} (\bibinfo {year} {1995})}\BibitemShut
  {NoStop}%
\bibitem [{\citenamefont {L{\"{u}}beck}\ \emph {et~al.}(1998)\citenamefont
  {L{\"{u}}beck}, \citenamefont {Schreckenberg},\ and\ \citenamefont
  {Usadel}}]{Lubeck1998}%
  \BibitemOpen
  \bibfield  {author} {\bibinfo {author} {\bibfnamefont {S.}~\bibnamefont
  {L{\"{u}}beck}}, \bibinfo {author} {\bibfnamefont {M.}~\bibnamefont
  {Schreckenberg}}, \ and\ \bibinfo {author} {\bibfnamefont {K.~D.}\
  \bibnamefont {Usadel}},\ }\href@noop {} {\bibfield  {journal} {\bibinfo
  {journal} {Phys.\ Rev.\ E}\ }\textbf {\bibinfo {volume} {57}},\ \bibinfo
  {pages} {1171} (\bibinfo {year} {1998})}\BibitemShut {NoStop}%
\bibitem [{\citenamefont {Chowdhury}\ \emph {et~al.}(1997)\citenamefont
  {Chowdhury}, \citenamefont {Ghosh}, \citenamefont {Majumdar}, \citenamefont
  {Sinha},\ and\ \citenamefont {Stinchcombe}}]{Chowdhury1997}%
  \BibitemOpen
  \bibfield  {author} {\bibinfo {author} {\bibfnamefont {D.}~\bibnamefont
  {Chowdhury}}, \bibinfo {author} {\bibfnamefont {K.}~\bibnamefont {Ghosh}},
  \bibinfo {author} {\bibfnamefont {A.}~\bibnamefont {Majumdar}}, \bibinfo
  {author} {\bibfnamefont {S.}~\bibnamefont {Sinha}}, \ and\ \bibinfo {author}
  {\bibfnamefont {R.~B.}\ \bibnamefont {Stinchcombe}},\ }\href {\doibase
  10.1016/S0378-4371(97)00365-8} {\bibfield  {journal} {\bibinfo  {journal}
  {Physica A}\ }\textbf {\bibinfo {volume} {246}},\ \bibinfo {pages} {471}
  (\bibinfo {year} {1997})}\BibitemShut {NoStop}%
\bibitem [{\citenamefont {Chowdhury}\ \emph {et~al.}(1998)\citenamefont
  {Chowdhury}, \citenamefont {Pasupathy},\ and\ \citenamefont
  {Sinha}}]{Chowdhury1998}%
  \BibitemOpen
  \bibfield  {author} {\bibinfo {author} {\bibfnamefont {D.}~\bibnamefont
  {Chowdhury}}, \bibinfo {author} {\bibfnamefont {A.}~\bibnamefont
  {Pasupathy}}, \ and\ \bibinfo {author} {\bibfnamefont {S.}~\bibnamefont
  {Sinha}},\ }\href {\doibase 10.1007/s100510050502} {\bibfield  {journal}
  {\bibinfo  {journal} {Eur.\ Phys.\ J.\ B}\ }\textbf {\bibinfo {volume} {5}},\
  \bibinfo {pages} {781} (\bibinfo {year} {1998})}\BibitemShut {NoStop}%
\bibitem [{\citenamefont {Roters}\ \emph {et~al.}(1999)\citenamefont {Roters},
  \citenamefont {L{\"{u}}beck},\ and\ \citenamefont {Usadel}}]{Roters1999}%
  \BibitemOpen
  \bibfield  {author} {\bibinfo {author} {\bibfnamefont {L.}~\bibnamefont
  {Roters}}, \bibinfo {author} {\bibfnamefont {S.}~\bibnamefont
  {L{\"{u}}beck}}, \ and\ \bibinfo {author} {\bibfnamefont {K.~D.}\
  \bibnamefont {Usadel}},\ }\href {\doibase 10.1103/PhysRevE.59.2672}
  {\bibfield  {journal} {\bibinfo  {journal} {Phys.\ Rev.\ E}\ }\textbf
  {\bibinfo {volume} {59}},\ \bibinfo {pages} {2672} (\bibinfo {year}
  {1999})}\BibitemShut {NoStop}%
\bibitem [{\citenamefont {Chowdhury}\ \emph {et~al.}(2000)\citenamefont
  {Chowdhury}, \citenamefont {Kert{\'{e}}sz}, \citenamefont {Nagel},
  \citenamefont {Santen},\ and\ \citenamefont
  {Schadschneider}}]{Chowdhury20002}%
  \BibitemOpen
  \bibfield  {author} {\bibinfo {author} {\bibfnamefont {D.}~\bibnamefont
  {Chowdhury}}, \bibinfo {author} {\bibfnamefont {J.}~\bibnamefont
  {Kert{\'{e}}sz}}, \bibinfo {author} {\bibfnamefont {K.}~\bibnamefont
  {Nagel}}, \bibinfo {author} {\bibfnamefont {L.}~\bibnamefont {Santen}}, \
  and\ \bibinfo {author} {\bibfnamefont {A.}~\bibnamefont {Schadschneider}},\
  }\href {\doibase 10.1103/PhysRevE.61.3270} {\bibfield  {journal} {\bibinfo
  {journal} {Phys.\ Rev.\ E}\ }\textbf {\bibinfo {volume} {61}},\ \bibinfo
  {pages} {3270} (\bibinfo {year} {2000})}\BibitemShut {NoStop}%
\bibitem [{\citenamefont {Roters}\ \emph {et~al.}(2000)\citenamefont {Roters},
  \citenamefont {L{\"{u}}beck},\ and\ \citenamefont {Usadel}}]{Roters2000}%
  \BibitemOpen
  \bibfield  {author} {\bibinfo {author} {\bibfnamefont {L.}~\bibnamefont
  {Roters}}, \bibinfo {author} {\bibfnamefont {S.}~\bibnamefont
  {L{\"{u}}beck}}, \ and\ \bibinfo {author} {\bibfnamefont {K.~D.}\
  \bibnamefont {Usadel}},\ }\href {\doibase 10.1103/PhysRevE.61.3272}
  {\bibfield  {journal} {\bibinfo  {journal} {Phys.\ Rev.\ E}\ }\textbf
  {\bibinfo {volume} {61}},\ \bibinfo {pages} {3272} (\bibinfo {year}
  {2000})}\BibitemShut {NoStop}%
\bibitem [{\citenamefont {Kerner}\ \emph {et~al.}(2002)\citenamefont {Kerner},
  \citenamefont {Klenov},\ and\ \citenamefont {Wolf}}]{Kerner2002}%
  \BibitemOpen
  \bibfield  {author} {\bibinfo {author} {\bibfnamefont {B.~S.}\ \bibnamefont
  {Kerner}}, \bibinfo {author} {\bibfnamefont {S.~L.}\ \bibnamefont {Klenov}},
  \ and\ \bibinfo {author} {\bibfnamefont {D.~E.}\ \bibnamefont {Wolf}},\
  }\href {\doibase 10.1088/0305-4470/35/47/303} {\bibfield  {journal} {\bibinfo
   {journal} {J.\ Phys.\ A-Math.\ Gen.}\ }\textbf {\bibinfo {volume} {35}},\
  \bibinfo {pages} {9971} (\bibinfo {year} {2002})}\BibitemShut {NoStop}%
\bibitem [{\citenamefont {Vilar}\ and\ \citenamefont
  {Souza}(1994)}]{Vilar1994}%
  \BibitemOpen
  \bibfield  {author} {\bibinfo {author} {\bibfnamefont {L.~C.~Q.}\
  \bibnamefont {Vilar}}\ and\ \bibinfo {author} {\bibfnamefont {A.~M.~C.}\
  \bibnamefont {Souza}},\ }\href@noop {} {\bibfield  {journal} {\bibinfo
  {journal} {Physica A}\ }\textbf {\bibinfo {volume} {211}},\ \bibinfo {pages}
  {84} (\bibinfo {year} {1994})}\BibitemShut {NoStop}%
\bibitem [{\citenamefont {Jost}\ and\ \citenamefont {Nagel}(2003)}]{Jost2003}%
  \BibitemOpen
  \bibfield  {author} {\bibinfo {author} {\bibfnamefont {D.}~\bibnamefont
  {Jost}}\ and\ \bibinfo {author} {\bibfnamefont {K.}~\bibnamefont {Nagel}},\
  }\href {\doibase 10.3141/1852-19} {\bibfield  {journal} {\bibinfo  {journal}
  {Transport Res.\ Rec.}\ }\textbf {\bibinfo {volume} {1852}},\ \bibinfo
  {pages} {152} (\bibinfo {year} {2003})}\BibitemShut {NoStop}%
\bibitem [{\citenamefont {Miedema}\ \emph {et~al.}(2014)\citenamefont
  {Miedema}, \citenamefont {de~Wijn},\ and\ \citenamefont
  {Schall}}]{Miedema2014}%
  \BibitemOpen
  \bibfield  {author} {\bibinfo {author} {\bibfnamefont {D.~M.}\ \bibnamefont
  {Miedema}}, \bibinfo {author} {\bibfnamefont {A.~S.}\ \bibnamefont
  {de~Wijn}}, \ and\ \bibinfo {author} {\bibfnamefont {P.}~\bibnamefont
  {Schall}},\ }\href {\doibase 10.1103/PhysRevE.89.062812} {\bibfield
  {journal} {\bibinfo  {journal} {Phys.\ Rev.\ E}\ }\textbf {\bibinfo {volume}
  {89}},\ \bibinfo {pages} {062812} (\bibinfo {year} {2014})}\BibitemShut
  {NoStop}%
\bibitem [{\citenamefont {Souza}\ and\ \citenamefont
  {Vilar}(2009)}]{Souza2009}%
  \BibitemOpen
  \bibfield  {author} {\bibinfo {author} {\bibfnamefont {A.~M.~C.}\
  \bibnamefont {Souza}}\ and\ \bibinfo {author} {\bibfnamefont {L.~C.~Q.}\
  \bibnamefont {Vilar}},\ }\href {\doibase 10.1103/PhysRevE.80.021105}
  {\bibfield  {journal} {\bibinfo  {journal} {Phys.\ Rev.\ E}\ }\textbf
  {\bibinfo {volume} {80}},\ \bibinfo {pages} {021105} (\bibinfo {year}
  {2009})}\BibitemShut {NoStop}%
\bibitem [{\citenamefont {Zhang}\ \emph {et~al.}(2011)\citenamefont {Zhang},
  \citenamefont {Zhang},\ and\ \citenamefont {Chen}}]{Zhang2011}%
  \BibitemOpen
  \bibfield  {author} {\bibinfo {author} {\bibfnamefont {W.}~\bibnamefont
  {Zhang}}, \bibinfo {author} {\bibfnamefont {W.}~\bibnamefont {Zhang}}, \ and\
  \bibinfo {author} {\bibfnamefont {W.}~\bibnamefont {Chen}},\ }\href
  {http://arxiv.org/abs/1102.5704} {}\bibinfo {howpublished} {e-print
  arXiv:1102.5704 [nlin.CG]} (\bibinfo {year} {2011})\BibitemShut {NoStop}%
\bibitem [{\citenamefont {Zhang}\ and\ \citenamefont
  {Zhang}(2014)}]{Zhang2014}%
  \BibitemOpen
  \bibfield  {author} {\bibinfo {author} {\bibfnamefont {W.}~\bibnamefont
  {Zhang}}\ and\ \bibinfo {author} {\bibfnamefont {W.}~\bibnamefont {Zhang}},\
  }\href {\doibase 10.1140/epjb/e2013-40506-4} {\bibfield  {journal} {\bibinfo
  {journal} {Eur.\ Phys.\ J.\ B}\ }\textbf {\bibinfo {volume} {87}},\ \bibinfo
  {pages} {4} (\bibinfo {year} {2014})}\BibitemShut {NoStop}%
\bibitem [{\citenamefont {de~Wijn}\ \emph {et~al.}(2012)\citenamefont
  {de~Wijn}, \citenamefont {Miedema}, \citenamefont {Nienhuis},\ and\
  \citenamefont {Schall}}]{DeWijn2012}%
  \BibitemOpen
  \bibfield  {author} {\bibinfo {author} {\bibfnamefont {A.~S.}\ \bibnamefont
  {de~Wijn}}, \bibinfo {author} {\bibfnamefont {D.~M.}\ \bibnamefont
  {Miedema}}, \bibinfo {author} {\bibfnamefont {B.}~\bibnamefont {Nienhuis}}, \
  and\ \bibinfo {author} {\bibfnamefont {P.}~\bibnamefont {Schall}},\ }\href
  {\doibase 10.1103/PhysRevLett.109.228001} {\bibfield  {journal} {\bibinfo
  {journal} {Phys.\ Rev.\ Lett.}\ }\textbf {\bibinfo {volume} {109}},\ \bibinfo
  {pages} {228001} (\bibinfo {year} {2012})}\BibitemShut {NoStop}%
\bibitem [{\citenamefont {Lakouari}\ \emph {et~al.}(2014)\citenamefont
  {Lakouari}, \citenamefont {Jetto}, \citenamefont {Ez-Zahraouy},\ and\
  \citenamefont {Benyoussef}}]{Lakouari2014}%
  \BibitemOpen
  \bibfield  {author} {\bibinfo {author} {\bibfnamefont {N.}~\bibnamefont
  {Lakouari}}, \bibinfo {author} {\bibfnamefont {K.}~\bibnamefont {Jetto}},
  \bibinfo {author} {\bibfnamefont {H.}~\bibnamefont {Ez-Zahraouy}}, \ and\
  \bibinfo {author} {\bibfnamefont {A.}~\bibnamefont {Benyoussef}},\ }\href
  {\doibase 10.1142/S0129183113500897} {\bibfield  {journal} {\bibinfo
  {journal} {Int.\ J.\ Mod.\ Phys.\ C}\ }\textbf {\bibinfo {volume} {25}},\
  \bibinfo {pages} {1350089} (\bibinfo {year} {2014})}\BibitemShut {NoStop}%
\bibitem [{\citenamefont {Kerner}(2013)}]{Kerner2013}%
  \BibitemOpen
  \bibfield  {author} {\bibinfo {author} {\bibfnamefont {B.~S.}\ \bibnamefont
  {Kerner}},\ }\href@noop {} {\bibfield  {journal} {\bibinfo  {journal}
  {Physica A}\ }\textbf {\bibinfo {volume} {392}},\ \bibinfo {pages} {5261}
  (\bibinfo {year} {2013})}\BibitemShut {NoStop}%
\bibitem [{\citenamefont {Kerner}\ \emph {et~al.}(2014)\citenamefont {Kerner},
  \citenamefont {Klenov},\ and\ \citenamefont {Schreckenberg}}]{Kerner3phase}%
  \BibitemOpen
  \bibfield  {author} {\bibinfo {author} {\bibfnamefont {B.~S.}\ \bibnamefont
  {Kerner}}, \bibinfo {author} {\bibfnamefont {S.~L.}\ \bibnamefont {Klenov}},
  \ and\ \bibinfo {author} {\bibfnamefont {M.}~\bibnamefont {Schreckenberg}},\
  }\href@noop {} {\bibfield  {journal} {\bibinfo  {journal} {Phys.\ Rev.\ E}\
  }\textbf {\bibinfo {volume} {89}},\ \bibinfo {pages} {052807} (\bibinfo
  {year} {2014})}\BibitemShut {NoStop}%
\bibitem [{Note1()}]{Note1}%
  \BibitemOpen
  \bibinfo {note} {The gap between the cars is $r-1$.}\BibitemShut {Stop}%
\bibitem [{\citenamefont {Gerwinski}\ and\ \citenamefont
  {Krug}(1999)}]{Gerwinski1999}%
  \BibitemOpen
  \bibfield  {author} {\bibinfo {author} {\bibfnamefont {M.}~\bibnamefont
  {Gerwinski}}\ and\ \bibinfo {author} {\bibfnamefont {J.}~\bibnamefont
  {Krug}},\ }\href@noop {} {\bibfield  {journal} {\bibinfo  {journal} {Phys.\
  Rev.\ E}\ }\textbf {\bibinfo {volume} {60}},\ \bibinfo {pages} {188}
  (\bibinfo {year} {1999})}\BibitemShut {NoStop}%
\bibitem [{\citenamefont {Fisher}(1967)}]{Fisher1967}%
  \BibitemOpen
  \bibfield  {author} {\bibinfo {author} {\bibfnamefont {M.~E.}\ \bibnamefont
  {Fisher}},\ }\href {\doibase 10.1088/0034-4885/30/2/306} {\bibfield
  {journal} {\bibinfo  {journal} {Rep.\ Prog.\ Phys.}\ }\textbf {\bibinfo
  {volume} {30}},\ \bibinfo {pages} {615} (\bibinfo {year} {1967})}\BibitemShut
  {NoStop}%
\bibitem [{\citenamefont {Barber}(1983)}]{Barber1983}%
  \BibitemOpen
  \bibfield  {author} {\bibinfo {author} {\bibfnamefont {M.~N.}\ \bibnamefont
  {Barber}},\ }in\ \href@noop {} {\emph {\bibinfo {booktitle} {Phase
  Transitions and Critical Phenomena}}}\ (\bibinfo  {publisher} {Academic
  Press, London},\ \bibinfo {year} {1983})\BibitemShut {NoStop}%
\bibitem [{\citenamefont {Ferrenberg}\ and\ \citenamefont
  {Landau}(1991)}]{Ferrenberg1991}%
  \BibitemOpen
  \bibfield  {author} {\bibinfo {author} {\bibfnamefont {A.~M.}\ \bibnamefont
  {Ferrenberg}}\ and\ \bibinfo {author} {\bibfnamefont {D.~P.}\ \bibnamefont
  {Landau}},\ }\href {\doibase 10.1103/PhysRevB.44.5081} {\bibfield  {journal}
  {\bibinfo  {journal} {Phys.\ Rev.\ B}\ }\textbf {\bibinfo {volume} {44}},\
  \bibinfo {pages} {5081} (\bibinfo {year} {1991})}\BibitemShut {NoStop}%
\bibitem [{\citenamefont {Balouchi}(2016)}]{Thesis}%
  \BibitemOpen
  \bibfield  {author} {\bibinfo {author} {\bibfnamefont {A.}~\bibnamefont
  {Balouchi}},\ }\href@noop {} {\bibinfo {type} {{Ph.D.} thesis}},\ \bibinfo
  {school} {Louisiana State University} (\bibinfo {year} {2016})\BibitemShut
  {NoStop}%
\bibitem [{Note2()}]{Note2}%
  \BibitemOpen
  \bibinfo {note} {The expression for $\chi _{4}$ here is the equal time
  correlation $\chi _{4}(0)$ used in Ref.~\cite {DeWijn2012}.}\BibitemShut
  {Stop}%
\end{thebibliography}%

\end{document}